\documentclass[aps,prd,nofootinbib,onecolumn,a4paper]{revtex4}
\usepackage{bm,latexsym,amsmath,amssymb,amsfonts,mathrsfs,amsfonts}
\usepackage[dvipdfmx]{graphicx}
\usepackage[hang,small,bf]{caption}
\usepackage[subrefformat=parens]{subcaption}
\usepackage{ascmac}
\usepackage{physics}
\usepackage{color,float}
\usepackage{braket}
\usepackage{bbm}
\usepackage{comment}

\newcommand{\simgt}{\lower.5ex\hbox{$\; \buildrel > \over \sim \;$}}
\newcommand{\simlt}{\lower.5ex\hbox{$\; \buildrel < \over \sim \;$}}
\def\barC{{\bar C}}

\begin{document}
\title{
Quantum state of a suspended mirror coupled to cavity light \\
		$-$ Wiener filter analysis of the pendulum and rotational modes $-$
}
	
	\author{Tomoya Shichijo}
        \email{shichijo.tomoya.351@kyudai.jp}
        \affiliation{Department of Physics, Kyushu University, 744 Motooka, Nishi-Ku, Fukuoka 819-0395, Japan}
	\author{Nobuyuki Matsumoto}
        \email{nobuyuki.matsumoto@gakushuin.ac.jp}
	\affiliation{Department of Physics, Faculty of Science, Gakushuin University, 1-5-1, Mejiro, Toshima, Tokyo, 171-8588 Japan}
	\author{Akira Matsumura}
        \email{matsumura.akira@phys.kyushu-u.ac.jp}
        \affiliation{Department of Physics, Kyushu University, 744 Motooka, Nishi-Ku, Fukuoka 819-0395, Japan}
        \author{Daisuke Miki}
        \email{miki.daisuke@phys.kyushu-u.ac.jp}
        \affiliation{Department of Physics, Kyushu University, 744 Motooka, Nishi-Ku, Fukuoka 819-0395, Japan}
	\author{Yuuki Sugiyama}
        \email{sugiyama.yuki@phys.kyushu-u.ac.jp}
        \affiliation{Department of Physics, Kyushu University, 744 Motooka, Nishi-Ku, Fukuoka 819-0395, Japan}
        \author{Kazuhiro Yamamoto}
        \email{yamamoto@phys.kyushu-u.ac.jp}
        \affiliation{Department of Physics, Kyushu University, 744 Motooka, Nishi-Ku, Fukuoka 819-0395, Japan}	
\affiliation{Research Center for Advanced Particle Physics, Kyushu University, 744 Motooka, Nishi-ku, Fukuoka 819-0395, Japan}
\affiliation{
International Center for Quantum-Field Measurement Systems for Studies of the Universe and Particles (QUP), KEK, Oho 1-1, Tsukuba, Ibaraki 305-0801, Japan}
\begin{abstract}
We investigated the quantum state of an optomechanical suspended mirror under continuous measurement and feedback control using Wiener filtering. We focus on the impact of the two-mode theory of a suspended mirror
on the quantum state, which is described by the pendulum and rotational modes. It is derived from the beam model coupled to the cavity light in the low-frequency regime, including the internal friction of the beam and the finite-size effect of the mirror. 
We constructed a Wiener filter for the two-mode theory and predicted the quantum state by evaluating the conditional covariance matrix using Wiener filter analysis. The results demonstrate that multimode analysis may play an important role in generating the quantum squeezed state. We also point out the possibility that one-mode analysis can be a good approximation by choosing the range of the Fourier space in the Wiener filter analysis properly.
\vspace{0cm}
\end{abstract}
\maketitle
\section{Introduction}
Optomechanical systems contribute significantly to the fundamental physics of gravitational waves, as well as macroscopic quantum systems and quantum sensing (e.g.,~\cite{Aspelmeyer, Bowen, Yambei, Michimura,Croquette}). Refs.~\cite{Korppi,Kotler,Lepinay} experimentally demonstrated that quantum entanglement occurs between nanoscale objects. These studies pioneered the boundary between quantum and classical worlds.
Gravity is a fundamental law of the classical world at the macroscopic scale and
the spacetime curvature is the gravity itself in general relativity.  
In recent years, the nature of gravity as a macroscopic quantum system has attracted considerable attention, whether it follows the framework of quantum mechanics \cite{Feynman}.
No one has verified the quantum nature of gravity; however, state-of-the-art quantum techniques may allow this in the future \cite{Tabletop,Bose,MV}. 
To verify the quantum nature of gravity, we must observe the gravitational interactions between objects in the quantum states. Therefore preparing the quantum state of massive objects is essential for the test because the gravitational force is very weak. The optomechanical oscillator system is a promising method for preparing the quantum states of massive objects as tabletop experiments \cite{Matsumoto, MY,Meng}, and many theoretical studies have focused on gravity-induced quantum entanglement generated between the gravitationally interacting mechanical oscillators \cite{Blaushi, Miao, Matsumura, Krisnanda, Datta, Miki2}.
To realize an optomechanical system in a quantum state, thermal fluctuations must be suppressed. Continuous measurement cooling technology has been developed to realize such a quantum state \cite{Matsumoto, MY, Meng}.
Constructing a precise theoretical model corresponding to an experimental setting will help investigate the feasibility of future experiments.
Ref.~\cite{Genes} provided a general framework for describing the cooling of a mechanical oscillator to the ground state using cavity detuning and feedback control.
Refs.~\cite{Vitali, Miao10} discussed the feasibility of detecting entanglement between the optical cavity modes and a mechanical oscillator in the ground state.

Ref.~\cite{MY} reported the mechanical squeezing of a milligram-scale suspended mirror with continuous measurement and feedback control using Wiener filtering \cite{Wiener} (also see \cite{Matsumoto, Schmole, Lopez}). 
A previous study~\cite{MY} focused only on the translational motion of the mirror.
The actual motion of the mirror comprises various motions such as rotation and string vibration.
Therefore, to obtain better measurements, the effects of these motions on the translational motion of the mirror must be considered.
In a recent study \cite{Sugiyama}, we investigated a theoretical model of 
a suspended mirror coupled to cavity light based on the beam model \cite{Saulson}.
We developed a theory for the optomechanical suspended mirror described by a two-mode system of pendulum and rotation modes in the low-frequency region, where the violin modes of the beam can be neglected. 
An important point is that the effect of the internal friction of the beam was explicitly considered using a complex Young's modulus. 
In the present paper, as an application of this previous work \cite{Sugiyama}, we investigate the quantum state of an optomechanical suspended mirror
described by the two-mode theory under continuous measurement and feedback control using Wiener filter analysis. 
We discuss how the mechanically squeezed quantum state of the mirror can be obtained when its rotational mode is considered.

The remainder of this study is organized as follows. 
In Sec.~\ref{beam}, we introduce a beam model coupled with cavity light.
The equations of motion are derived from the Hamiltonian of the system comprising a mirror described by the position of the center of mass and the rotation around it, which are coupled to the beam and cavity light. The equations of motion were solved using a perturbative method around a steady-state solution.
After briefly reviewing the theoretical framework of the model in subsection \ref{beamA}, we derive the equations using the perturbation approach in subsection \ref{secpert} and the steady-state solution in subsection \ref{beamB}.
We show that the perturbation equations are reduced to those of the two-mode theory of the pendulum and rotational modes in the low-frequency regime in subsection \ref{beamC}. 
In subsection \ref{beamD}, we discuss the validity of the structural damping model based on a beam model with a complex Young's modulus.
Assuming the model parameters of an experiment in the near future, we find the solution for the equations including the dissipation and noise to describe the model with feedback cooling in the perturbation equations in subsection \ref{beamE}.
In Sec.~\ref{wiener}, we evaluated the conditional covariance with respect to the pendulum and rotational modes using the Wiener filter process based on the two-mode theory.
In Sec.~\ref{secdisc}, we discuss the validity of the possible use of the one-mode filter with the constrained range in Fourier space, where we also discuss an implication of the present investigation for generating entanglement between mirrors demonstrated in Ref.~\cite{Miki3}.
Sec.~\ref{summary} presents the summary and conclusions.
In Appendix~\ref{appFilter}, we explained the details of finding the Wiener filter 
of the two-mode theory in Sec.~\ref{wiener}. 
In Appendix~\ref{point}, we present a Wiener filter of the one-mode theory in which the mirror is treated as a point particle with only the pendulum mode. We use it to evaluate the conditional variances of the pendulum mode (see Ref.~\cite{MY}) for comparison with the two-mode model.
\def\x{{q}}
\def\A{{x}}
\def\B{{y}}
\begin{figure}[b]
\centering
\includegraphics[width=0.7\linewidth]{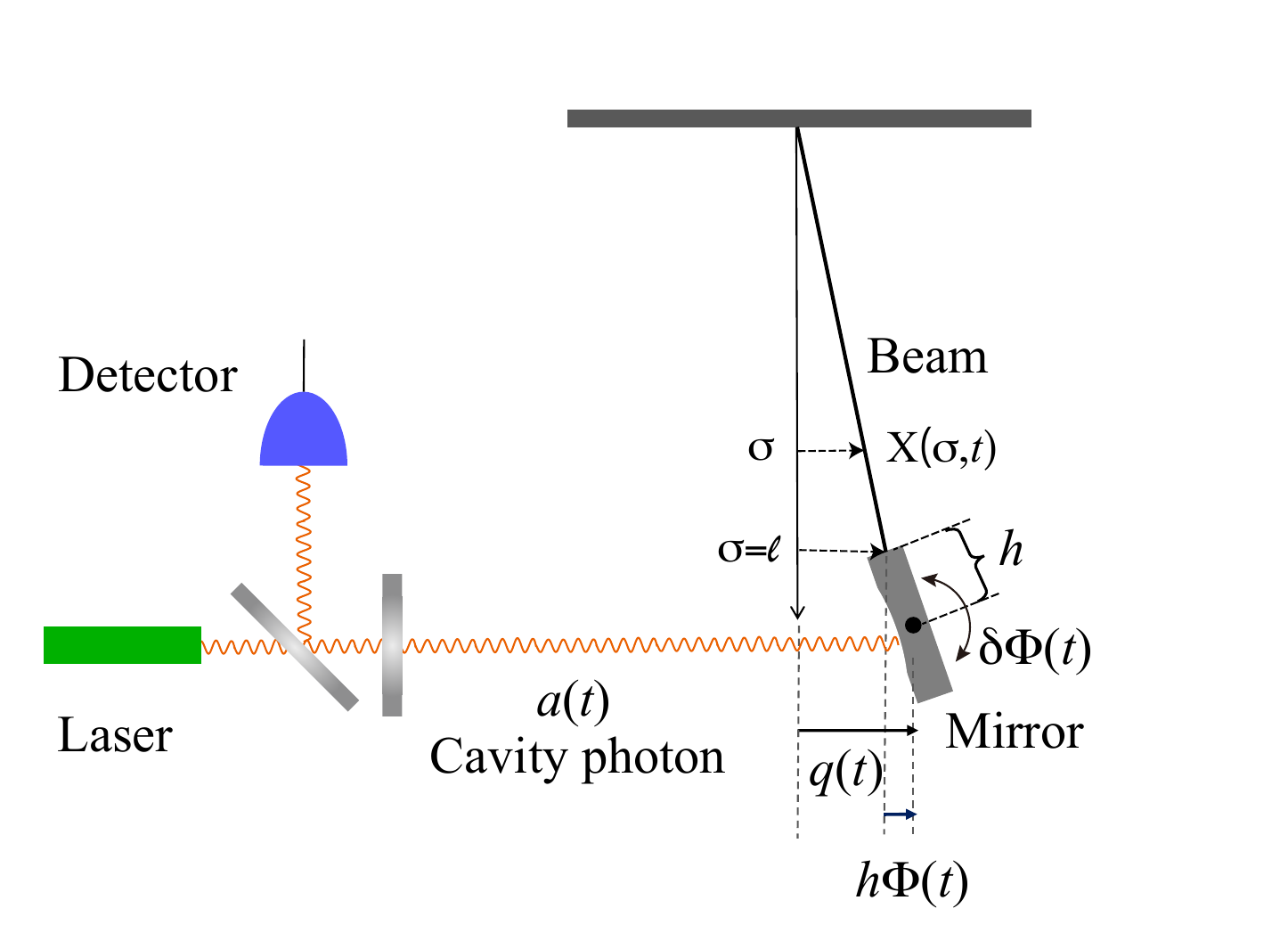}
\caption{Schematic plot of the beam model coupled to cavity light.}
\label{fig:beamf1}
\end{figure}

\section{Optomechanical Beam model\label{beam}}
\subsection{Basic formulation \label{beamA}}
In this section, we introduce the beam model \cite{Saulson} coupled to cavity photons (see Ref.~\cite{Sugiyama}). 
The beam model describes the motion of the wire that suspends the mirror as a finite-sized object (see Figure \ref{fig:beamf1}).
We assume that the mirror is a rigid body with a translational
motion as a pendulum and rotational motion about its center of mass.
In the beam model, the motion of the mirror is characterized in connection with the dynamics of the beam. We assume that the beam has a complex Young's modulus $E(=E_0 (1-i \phi))$, area moment of inertia $I$, and mass per unit length $\rho$.
One end of the beam is completely fixed and the mirror is fixed at the other end, separated by length $\ell$, with a distance $h$ from the center of mass.
The mass of mirror $M$ has a moment of inertia $J$ around its center of mass.
The tension of the beam is $T=M g$ because of the balance of the gravitational force on the mass.
The transverse mode of the beam is expressed as $X(\sigma,t)$ in the range $0 \leq \sigma \leq \ell$.
In the beam model, we assume that the position of the center of mass of the mirror $\x(t)$ is described by the following boundary conditions for the beam:
\begin{eqnarray}
\Phi (t) &=&\left. \frac{\partial X(\sigma,t)}{\partial \sigma} \right|_{\sigma=\ell} ,   \label{condition_Phi} \\
\x(t)&=&X(\sigma,t)\vert_{\sigma=\ell}+h \Phi (t) \label{condition_x}, 
\end{eqnarray}
and 
\begin{eqnarray}
X({\sigma=0,t})&=&\left. \frac{\partial X(\sigma,t)}{\partial \sigma} \right|_{\sigma=0}=0, \label{condition_X}
\end{eqnarray}
where $\Phi$ is the tilt of the beam at the lower end with respect to the vertical direction, which we call the rotational mode of the mirror.
 The physical meaning of the boundary conditions is that the beam is perpendicularly connected to the ceiling and the mirror at both the ends $\sigma=0$ and $\sigma=\ell$, respectively.
We note that the boundary condition
under the above assumptions, the Lagrangian of the beam model can be written as
\begin{eqnarray}
\mathcal{L}_B=K-V_1 -V_2,
\end{eqnarray}
where $K$ is the kinetic energy of the mirror and beam in the transverse and rotational modes, $V_{1}$ is the potential energy due to gravity, and $V_2$ is the elastic energy of the beam, which is defined as
\begin{eqnarray}
K&\equiv&\frac{1}{2} \int_{0}^{\ell}d\sigma \rho \left(\frac{\partial X(\sigma,t)}{\partial t} \right)^2+\frac{1}{2}M \left(\frac{d \x(t)}{dt} \right)^2+\frac{1}{2}J \left(\frac{d \Phi(t)}{dt} \right)^2,
\\
V_1&\equiv&\frac{1}{2} \int_{0}^{\ell}d\sigma T \left(\frac{\partial X (\sigma,t)}{\partial \sigma} \right)^2+\frac{1}{2} T h\Phi^2(t),
\label{V1}
\\
V_2&\equiv&\frac{1}{2} \int_{0}^{\ell}d\sigma EI \left(\frac{\partial^2 X(\sigma,t)}{\partial \sigma^2}\right)^2.
\end{eqnarray}
 We note that the gravitational potential energy of the mirror is included in the second term of $V_1$ in Eq.~(\ref{V1}), $\frac{1}{2} T h\Phi^2(t)$, with $T=Mg$ under the small angle approximation $\Phi\ll1$. We do not take into account the gravitational energy of the beam because it is small within the same approximation and the mass of the beam is negligible.
The conjugate momentum with respect to the position of the beam and mirror $X$, $\x$, and the rotational mode $\Phi$ are defined as:
\begin{eqnarray}
\Pi(\sigma,t)& \equiv &\frac{\partial \mathcal{L}_B}{\partial \dot{X}(\sigma,t)}=\rho \dot{X}(\sigma,t), 
\\
p(t)&\equiv &\frac{\partial \mathcal{L}_B}{\partial \dot{\x}(t)}=M\dot{\x}(t), 
\\
\Pi_{_\Phi}(t)& \equiv &\frac{\partial \mathcal{L}_B}{\partial \dot{\Phi}(t)}=J \dot{\Phi}(t), 
\end{eqnarray}
where the dot denotes differentiation with respect to time.
Subsequently, we obtain the  Hamiltonian $\mathcal{H}_B$ of the beam model as follows:
\begin{eqnarray}
		\mathcal{H}_B= \int_{0}^{\ell}d\sigma \left\{ \frac{1}{2 \rho} \Pi^2(\sigma,t)+\frac{T}{2} \left(\frac{\partial X(\sigma,t)}{\partial \sigma}\right)^2+\frac{EI}{2}\left(\frac{\partial^2 X(\sigma,t)}{\partial \sigma^2}\right)^2\right\}
		+\frac{1}{2M} p^2(t)+\frac{1}{2J}\Pi_{\Phi}^2(t) +\frac{hT}{2}\Phi^2(t).
	\end{eqnarray}
	
	As we investigate the quantum state of an optomechanical suspended mirror under continuous measurement and feedback control, 
	we first include the dynamics of the cavity photon that is injected by a laser and trapped between the mirror suspended by the beam and another fixed mirror.
The interaction between a moving mirror and cavity photons was investigated in Ref.~\cite{Law}, in which the author derived the Hamiltonian directly from the equation of motion of a moving mirror and the wave equation of vector potential with time-varying boundary conditions. 
Following the result, the Hamiltonian $\mathcal{H}_C$ for the cavity photons is written as 
	\begin{eqnarray}
		\mathcal{H}_C=-\hbar G_0 a'^{\dag}a' \x+\hbar \omega_c a'^{\dag}a'+i \hbar \mathcal{E}( a'^{\dag} e^{-i \omega_0 t}-a' e^{i \omega_0 t}), \label{hamiltonian_c}
	\end{eqnarray}
 under the assumption that the two-photon emission and absorption can be neglected, where the first term of the right-hand side of this Hamiltonian describes the interaction between the mirror and the cavity photons, and  $G_0(\propto \omega_c/L)$ is the coupling constant between them, $\omega_c$ is the cavity photon frequency, $L$ is the cavity length, and $\hbar$ is the reduced Planck constant.
 The second term of Eq.~(\ref{hamiltonian_c}) is the Hamiltonian of the free field of the cavity photon, while the last term describes the input laser at frequency $\omega_0$. $\mathcal{E}$ and the input laser power $P$ are related by $|\mathcal{E}|= \sqrt{2 P \kappa/\hbar \omega_0}$~\cite{Genes}, where $\kappa$ is the decay rate describing the rate of escape of photons from the cavity. 
 We assume that the mirror scatters the cavity photons to form the standing wave at all times. This assumption should be validated carefully. However, the parameters adopted in the present paper are based on the realistic experiment reported in Ref.~\cite{Matsumoto,MY}, which may allow us to assume our theoretical model forms a standing wave cavity mode.
 Also, the finesse of our experimental system is only a few thousand finesse, so the effect of optical loss is negligible.

	From the Hamiltonians $\mathcal{H}_{B}$ and $\mathcal{H}_{C}$, the equations of motion for the entire system are obtained as follows:
	\begin{eqnarray}
		\rho \ddot{X}
		&=&
		T\frac{\partial^2 X}{\partial \sigma^2}-EI \frac{\partial^4 X}{\partial \sigma^4}, \label{eom_X_or} 
		\\
		M \ddot{\x}&=&-T\Phi+EI \left. \frac{\partial^3 X}{\partial \sigma^3}\right|_{\sigma=\ell}+\hbar G_0 a^{\dag} a, \label{eom_x_or} 
		\\
		J \ddot{\Phi}&=&-EI\left(\left.\frac{\partial^2 X}{\partial \sigma^2}\right|_{\sigma=\ell}+h \left. \frac{\partial^3 X}{\partial \sigma^3}\right|_{\sigma=\ell} \right), \label{eom_phi_or} 
		\\
		\dot{a}&=&-\kappa a+i G_0 a \x-i (\omega_c- \omega_0) a +\mathcal{E}+\sqrt{2 \kappa}a_{\mathrm{in}}, \label{eom_a}
	\end{eqnarray}
	where $a=a' e^{i \omega_0 t}$ is introduced into Eq.~\eqref{eom_a} instead of $a'$.
	The noise term $\sqrt{2\kappa} a_{\rm in}$ is
	introduced in Eq.~\eqref{eom_a} to satisfy the fluctuation-dissipation relation. 
	Specifically, we assume that the decay rate $\kappa$ is related to the radiation input noise, 
	which is characterized by the correlation function of $a_{\text{in}}(t)$
	\begin{eqnarray}
		\langle a_{\text{in}}(t) a_{\text{in}}^\dag(t') \rangle&=&[N(\omega_c)+1]\delta(t-t'), 
		\\ 
		\langle a_{\text{in}}^{\dag}(t)  a_{\text{in}}(t') \rangle&=&N(\omega_c) \delta(t-t'). 
	\end{eqnarray} 
	Here $N(\omega_c)=(\mathrm{exp}\{\hbar \omega_c/k_B T_0\}-1)^{-1}$ denotes the average number of photons in the equilibrium state.
	Because the energy of the cavity photon $\hbar \omega_{\text{c}}$ is sufficiently large compared to the room temperature $k_{\text{B}}T_{0}$: $\hbar \omega_c/k_B T_0 \gg 1$, it can be regarded as $N(\omega_{\text{c}}) \simeq 0$.
In the latter part of the present paper, we consider the Langevin equation by including the noise and dissipation terms for the equations around the steady-state solutions. 

\subsection{Perturbative approach}
\label{secpert}
	Equations (\ref{eom_X_or})$\sim$(\ref{eom_a}) are decomposed into the steady-state part and its perturbed part by setting
	\begin{eqnarray}
		X&=&\bar{X}(\sigma)+\delta X(\sigma,t),  \\
		\x&=&\bar{\x}+\delta \x(t), \\
		\Phi&=&\bar{\Phi}+\delta \Phi(t),  \\
		a&=&\bar{a}+\delta a(t). 
	\end{eqnarray}
	The boundary conditions Eqs.~(\ref{condition_x}) $\sim$ (\ref{condition_X})
	lead to the following conditions for the steady-state solution:
	\begin{eqnarray}
		\bar{\x}&=&\bar{X}(\sigma=\ell)+h \bar{\Phi}, \label{bg_cnd_Q} \\
		\bar{\Phi}&=&\left. \frac{\partial \bar{X}(\sigma)}{\partial \sigma} \right|_{\sigma=\ell}, \label{bg_cnd_Phi} \\
		\bar{X}(\sigma=0)&=&\left. \frac{\partial \bar{X}}{\partial \sigma}\right|_{\sigma=0}=0. \label{bg_cnd_0}
	\end{eqnarray}
	Because the steady state is invariant with respect to time, its equations of motion become
	\begin{eqnarray}
		&&T\frac{\partial^2 \bar{X}}{\partial \sigma^2}-EI \frac{\partial^4 \bar{X}}{\partial \sigma^4}=0, \label{bg_X} \\
		&&-T\bar{\Phi}+EI \left. \frac{\partial^3  \bar{X}}{\partial \sigma^3}\right|_{\sigma=\ell} +\hbar G_0 |\bar{a}|^2=0, \label{bg_x} \\
		&&-EI \left( \left. \frac{\partial^2 \bar{X}}{\partial \sigma^2}\right|_{\sigma=\ell}+h \left. \frac{\partial^3 \bar{X}}{\partial \sigma^3}\right|_{\sigma=\ell} \right)=0, \label{bg_Phi} \\
		&&-\left\{\kappa+i(\omega_c-\omega_0-G_0 \bar{\x}) \right\} \bar{a}+\mathcal{E}=0. \label{bg_a}
	\end{eqnarray}
	
	On the other hand, the boundary conditions for the perturbed part are also obtained as
	\begin{eqnarray}
		\delta \x(t)&=&\delta X(\sigma=\ell,t)+h \delta \Phi(t),
		\\
		\delta \Phi(t)&=&\left. \frac{\partial \delta X(\sigma,t)}{\partial \sigma} \right|_{\sigma=\ell},
		\\
		\delta X(\sigma=0,t)&=&\left. \frac{\partial \delta X(\sigma,t)}{\partial \sigma}\right|_{\sigma=0}=0.
	\end{eqnarray}
	The equations of motion for the perturbed part are as follows:
	\begin{eqnarray}
		\rho \delta \ddot{X}&=&T\frac{\partial^2 \delta X}{\partial \sigma^2}-EI\frac{\partial^4 \delta X}{\partial \sigma^4}, \\
		M \delta \ddot{\x}&=&-T \delta\Phi +E I\left. \frac{\partial^3 \delta X}{\partial \sigma^3}\right|_{\sigma=\ell}+\hbar G_0 (\bar{a}\delta a+\bar{a}^{*}\delta a^{\dag}), \label{1st_q}
		\\
		J \delta \ddot{\Phi}&=&-EI \left(\left. \frac{\partial^2 \delta X}{\partial \sigma^2}\right|_{\sigma=\ell} +h\left. \frac{\partial^3 \delta X}{\partial \sigma^3}\right|_{\sigma=\ell} \right), \\
		\delta \dot{a}&=&-\{\kappa +i (\omega_c-\omega_0)\} \delta a+i G_0(\bar{a}\delta \x+\bar{\x}\delta a)+\sqrt{2 \kappa}a_{\mathrm{in}}. \label{1st_eom_a}
	\end{eqnarray}
These are the linearized equations around the steady-state solution. We note that the terms with $G_0$ in Eqs.~(\ref{1st_q}) and (\ref{1st_eom_a}) come from  the radiation pressure interaction linearized around the steady-state solution.

	\subsection{Steady-state solution\label{beamB}}
	In this section, we obtain the steady-state solution.
First, from Eq.~(\ref{bg_X}), we easily have
 \begin{eqnarray}
  {\partial^2 \bar{X}(\sigma)\over\partial \sigma^2}=A''e^{\beta\sigma}+B''e^{-\beta\sigma},
 \end{eqnarray}
 where we define $\beta \equiv \sqrt{{T}/{EI}}$ and $A''$ and $B''$ are integration constants. The second-order integration leads to the general solution for the beam displacement $\bar{X}(\sigma)$ is expressed as follows:
	\begin{eqnarray}
		\bar{X}(\sigma)=A' e^{\beta \sigma}+B' e^{-\beta \sigma} +C' \sigma+D',
  \label{barX_0}
	\end{eqnarray}
	where $A', B', C', D'$ are constants. 
	Using boundary condition Eq.~(\ref{bg_cnd_0}), we obtain the following conditions:
	\begin{eqnarray}
		C'=-\beta (A'-B'),
		\quad
		D'=-(A'+B') \label{D'}.
		\label{CaD}
	\end{eqnarray}
	Substituting Eqs.~(\ref{barX_0}),(\ref{CaD}) into Eq.~(\ref{bg_Phi}) yields
	\begin{eqnarray}
		0=A' \beta^2 e^{\beta \ell}+B' \beta^2 e^{-\beta \ell}+h \beta^3 (A' e^{\beta \ell}-B' e^{-\beta \ell}), 
	\end{eqnarray}
	which is rewritten as
	\begin{eqnarray}
		B'=-\frac{1+h \beta}{1-h \beta}e^{2\beta \ell}A'.
	\end{eqnarray}
	In addition, substituting Eqs.~(\ref{barX_0}), (\ref{CaD}) into Eq.~(\ref{bg_x}) yields
	\begin{eqnarray}
		C'&=&\frac{\hbar G_0 |\bar{a}|^2}{T}.
	\end{eqnarray}
	Substituting these relationships into Eq.~(\ref{D'}), we obtain
	\begin{eqnarray}
		A'= -\frac{\hbar G_0 |\bar{a}|^2}{\beta T} \frac{1-h \beta}{(1+h \beta) e^{2 \beta \ell}+1-h \beta}, 
		\quad
		D'= A'\left(\frac{1+h \beta}{1-h \beta} e^{2\beta \ell}-1 \right).
	\end{eqnarray}
	Finally, we obtain the steady-state solution of the beam 
	displacement $\bar{X}(\sigma)$ 
	\begin{eqnarray}
		\bar{X}(\sigma)&=&-\frac{\hbar G_0 |\bar{a}|^2}{\beta T} \frac{1-h \beta}{(1+h \beta) e^{2 \beta \ell}+1-h \beta}
		\left[e^{\beta \ell}+\frac{1+h \beta}{1-h \beta}e^{2 \beta \ell}(1-e^{-\beta \sigma})-1
		\right]+\frac{\hbar G_0 |\bar{a}|^2}{T} \sigma \nonumber \\
		&&\simeq \frac{\hbar G_0 |\bar{a}|^2}{T}  \sigma,
	\end{eqnarray}
	where the second line represents an approximate formula using $\beta \ell \gg 1$, 
    which is satisfied in our model. This condition means the smallness of the flexural rigidity of the beam.
	
\begin{figure}[t]
		\centering
		\includegraphics[width=100mm]{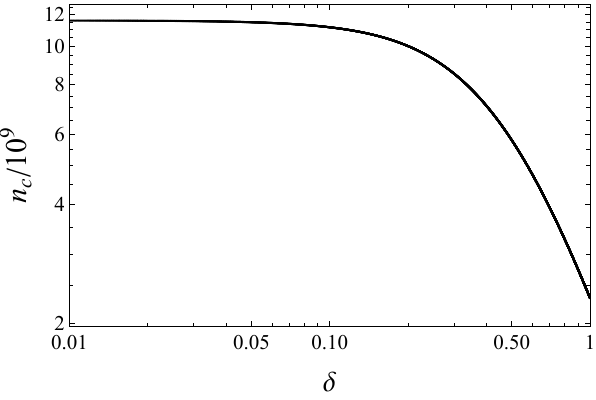}
		\caption{
			Number of photons $n_c=|\bar a|^2$ as a function of the normalized detuning~$\delta$($=-\Delta/2\kappa$). Here we adopted the parameters listed in 
			Table \ref{tab:parameter}.
			\label{fig:nc}}
		\vspace{1cm}
		\centering
		\includegraphics[width=100mm]{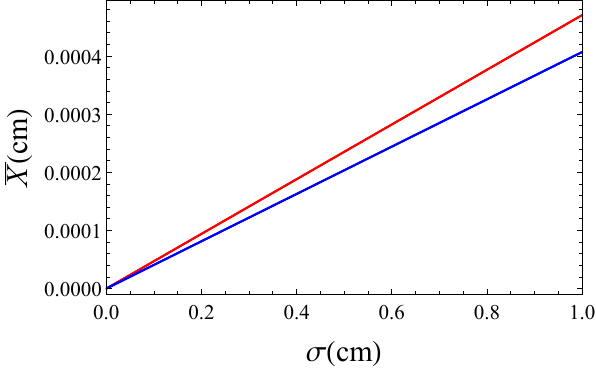}
		\caption
		{ Steady-state solution of the beam $\bar{X}(\sigma)$.
			The red and blue lines are plotted with $\delta=0.0292$ and $\delta=0.2$, respectively.
			\label{fig:barX}}
	\end{figure}

	From Eq.~(\ref{bg_cnd_Q}), the position of the center of mass $\bar{q}$ can be written as
	\begin{eqnarray}
		\bar{\x}&=&\frac{\hbar G_0 |\bar{a}|^2}{\beta T} \frac{1-h \beta}{(1+h \beta) e^{2 \beta \ell}+1-h \beta}
		\left[1+(h+\ell) \beta-\frac{1+h \beta}{1-h \beta} (1-\beta(h+\ell)) e^{2\beta \ell}\right]\nonumber \\
		&&\simeq \frac{\hbar G_0}{T}(\ell+h) |\bar{a}|^2,
	\end{eqnarray}
	where the second line is obtained by using $\beta \ell \gg 1$.
	From Eq.~(\ref{bg_a}), $\bar{a}$ can be written as
	\begin{eqnarray}
		\bar{a}=\frac{\mathcal{E}}{\kappa+i(\omega_c-\omega_0-G_0\bar{\x})}\equiv\alpha_{\text{s}}e^{-i\theta},
	\end{eqnarray}
	where the amplitude of the cavity photon $\bar{a}$ is a complex number, and we  
	introduced the phase $\theta$ and the positive and real number $\alpha_s$,
	which leads to 
	\begin{eqnarray}
		|\bar{a}|^2
		=
		\alpha^2_{\text{s}}
		=\frac{\mathcal{E}^2}{\kappa^2+(\omega_c-\omega_0-G_0 \bar{q})^2} =\frac{\mathcal{E}^2}{\kappa^2+\Delta^2}.
	\end{eqnarray}
	Here we introduced $\Delta=\omega_c-\omega_0-G_0 \bar{q}$ in the last equality.
	The number of photons in the cavity is denoted as $n_c=|\bar{a}|^2$.
	The detuning $\Delta$ is a parameter that can be experimentally adjusted by changing the cavity length $L$.
	It is normalized by the optical decay rate $\kappa$ and the opposite sign is defined as $\delta \equiv -\Delta/2\kappa$.
	In this case, the photon number $n_c$ can be expressed as a function of $\delta$ in the steady state as follows:
	\begin{eqnarray}
		n_c=|\bar a|^2=\frac{\mathcal{E}^2}{\kappa^2(1+\delta^2)}.
	\end{eqnarray}
	The analysis in the present study uses the values presented in Table~\ref{tab:parameter}, which were employed in Refs.~\cite{MY,Lopez} and are expected from state-of-the-art techniques (see Refs.~\cite{Miki3,Sugiyama}).
	Figure~\ref{fig:nc} shows the number of photons as a function of $\delta$. 
	Figure~\ref{fig:barX} shows  $\bar{X}(\sigma)$ in the steady state,
	which is proportional to $\sigma$. The red and blue curves represent $\delta=0.0292$ and $\delta=0.2$, respectively.
	\begin{table}[t]
		\centering
		\begin{tabular}{|c|c|c|c|c|}
			\hline
			Symbol & Description & Value & Dimension & Reference\\
			\hline
			$P$ &Laser Power & $30$ & mW & \cite{MY,Lopez}\\
			$\kappa/2\pi$ &Optical decay rate & $8.2 \times 10^{5}$ & $\mathrm{~Hz}$ & \cite{MY,Lopez} \\
			$\omega_{\text{c}}/2\pi \sim \omega_{0}/2\pi$ & Cavity resonance frequency & $2.818\times10^{14}$ &  $\mathrm{~Hz}$ & \cite{MY,Lopez}\\
			$\delta_0(=-\Delta_0/2\kappa)$ & Normalized  detuning & $0.2$ &  & \cite{Miki3}\\
			\hline
			$M$ & Mirror mass & $7.71 \times 10^{-3}$ & $\mathrm{~g}$ & \cite{MY,Lopez}\\
			$L$ & Cavity length & 9.81 & $\mathrm{cm}$ & \cite{MY,Lopez}\\
			$\ell$ & Length of beam & 1 & $\mathrm{~cm}$ & \cite{MY,Lopez}\\
			$h$ & Distance from center of mass to fixed point & 0.15 & $\mathrm{~cm}$ & \cite{MY,Lopez} \\
			$\Omega/2\pi$ & Frequency of Pendulum & $ 4.99$ & $\mathrm{~Hz}$ 
			& \cite{MY,Lopez} \\
			$J$ & Moment of inertia & $4.50 \times 10^{-5}$ & $\mathrm{~g} \mathrm{~cm}^{2}$ & \cite{Sugiyama} \\
			$E_{0} I$ & Flexural rigidity & $3.583 \times 10^{-6}$ & $\mathrm{g} \mathrm{~cm}^{3} / \mathrm{s}^{2}$ & \cite{Sugiyama}\\
			$\phi_0$ & Internal loss factor & $ 10^{-3}$ & & \cite{Sugiyama}\\
			$\rho$ & Mass per unit length & $1.72 \times 10^{-8}$ & $\mathrm{~g} / \mathrm{cm}$ & \cite{Sugiyama}\\
			$G_0/2\pi$ &Optomechanical coupling & $ 4.639 \times 10^{13}$ & $\mathrm{~Hz} / \mathrm{cm}$ & \cite{MY,Lopez} \\
			\hline
			$T_0$ &Bath Temperature & $300$ & $\mathrm{~K}$ & \cite{MY,Lopez} \\
			$\Gamma(\Omega)/2\pi$ & Mechanical decay rate without rotational mode
			& $4.11\times  10^{-7}$ & $\mathrm{~Hz}$ & \cite{Matsumoto} \\
   $\Gamma_{\text{r}}(\Omega)/2\pi$ & Mechanical decay rate including rotational mode
   & $1.717\times  10^{-6}$ & $\mathrm{~Hz}$ &  \\
			$\gamma_m/2\pi$ & Effective mechanical decay rate under feedback & $ 6.875 \times 10^{-3}$ & $\mathrm{~Hz}$ & \cite{Miki3} \\
			$N(\omega_c)$ & Equilibrium mean thermal photon number & $0 $ & $ $ & \cite{Miki3}\\
			\hline
			$g$ & Gravitational acceleration & 980 & $\mathrm{~cm} / \mathrm{s}^{2}$ & \\
			$c$ & Speed of light & $2.998 \times 10^{10}$ & $\mathrm{~cm} / \mathrm{s}$ &\\
			$\hbar$ & Reduced Planck constant & $1.05\times10^{-27}$ & $\mathrm{~g} \mathrm{~cm^2}/\mathrm{s}$ &\\
			$k_{\text{B}}$ & Boltzmann constant & $1.38\times 10^{-16}$ & $\mathrm{~g} \mathrm{~cm^2} / \mathrm{K \cdot s^2}$ &\\
			\hline
		\end{tabular}
		\caption{Parameters used in the present study.
		 	\label{tab:parameter}}
	\end{table}
	
	\subsection{Perturbation equations \label{beamC}}
	Next, we focus on the perturbation equations.
	For convenience, we introduce the amplitude and phase quadratures with respect to the cavity photon as
	\begin{eqnarray}
		\delta \A \equiv 
  \delta a+
  \delta a^{\dag},
		\quad
		\delta \B\equiv\frac{1}{i}(
  \delta a-
  \delta a^{\dag}), 
	\end{eqnarray}
	and the input noise as
	\begin{eqnarray}
		\A_{\mathrm{in}}\equiv 
 a_{\mathrm{in}}+ 
 a_{\mathrm{in}}^{\dag},
		\quad
		\B_{\mathrm{in}}\equiv\frac{1}{i} (
  a_{\mathrm{in}}- 
  a_{\mathrm{in}}^{\dag}).
	\end{eqnarray}
	From Eq.~(\ref{1st_eom_a}), $\delta x$ and $\delta y$ can be rewritten as
	\begin{eqnarray}
		\delta \dot{\A}
		=-\kappa \delta \A+\Delta \delta \B+\sqrt{2\kappa} \A_{\mathrm{in}},\quad
		\nonumber
		\delta \dot{\B}=-\kappa \delta \B-\Delta \delta \A +2 G_0 \sqrt{n_c} \delta \x+\sqrt{2 \kappa} \B_{\mathrm{in}}.
	\end{eqnarray}
	The perturbed part of the equations is solved in the Fourier space.
	In this study, the Fourier transform of $\delta f(t)$ is defined as:
	\begin{eqnarray}
		\delta f(\omega)=\int_{-\infty}^{\infty} \delta f(t) e^{i \omega t} dt,
	\end{eqnarray}
	where $\delta f$ denotes $\delta q$, $\delta \Phi,\delta X, \delta x, \delta y$. 
	Using this definition, the boundary conditions in the Fourier space are 
	\begin{eqnarray}
		&&\delta \x(\omega)=\delta X(\sigma=\ell,\omega)+h \delta \Phi(\omega), \label{bd_dx}  \\
		&&\delta \Phi(\omega)=\left. \frac{\partial \delta X(\sigma,\omega)}{\partial \sigma} \right|_{\sigma=\ell}, \label{bd_dP} \\
		&&\delta X (\sigma=0,\omega)=\left. \frac{\partial \delta X(\sigma,\omega)}{\partial \sigma}\right|_{\sigma=0}=0. \label{bd_dxP_0}
	\end{eqnarray}
	The perturbation equations are given by
	\begin{align}
		- \omega^2 \rho \delta X(\sigma,\omega)
		&=
		T\frac{\partial^2 \delta X(\sigma, \omega)}{\partial \sigma^2}-EI\frac{\partial^4 \delta X(\sigma,\omega)}{\partial \sigma^4}, \label{1st_X} 
		\\
		- \omega^2 M \delta \x(\omega)
		&=
		-T \delta\Phi(\omega) +E I\left. \frac{\partial^3 \delta X(\sigma, \omega)}{\partial \sigma^3}\right|_{\sigma=\ell}+ \hbar G_0 \sqrt{n_c}  \delta \A(\omega), \label{1st_x} 
		\\
		- \omega^2\ J \delta \Phi(\omega)
		&=
		-EI \left(\left. \frac{\partial^2 \delta X(\sigma, \omega)}{\partial \sigma^2}\right|_{\sigma=\ell} +h\left. \frac{\partial^3 \delta X(\sigma,\omega)}{\partial \sigma^3}\right|_{\sigma=\ell} \right), \label{1st_phi} 
		\\
		-i \omega \delta \A(\omega)
		&=
		-\kappa \delta \A(\omega)+\Delta \delta \B(\omega) +\sqrt{2 \kappa}\A_{\mathrm{in}}(\omega), \label{1st_A} 
		\\
		-i \omega \delta \B(\omega)&=
		-\kappa \delta \B(\omega)-\Delta \delta \A(\omega)+2 G_0 \sqrt{n_c} \delta \x(\omega) +\sqrt{2 \kappa} \B_{\mathrm{in}}(\omega). \label{1st_B}
	\end{align} 
Here we summarize the numbers of physical degrees of freedom of the system. The beam described by $\delta X(\sigma,\omega)$ has infinite numbers of degrees of freedom. 
The mirror is described by the position as the center of mass of the mirror, $\delta q(\omega)$, and the rotational mode $\delta \Phi(\omega)$, while the cavity photon is described by the amplitude quadrature and the phase quadrature $\delta x(\omega)$ and $\delta y(\omega)$. Thus, the total system consists of a one-dimensional field's dynamical degrees of freedom of the beam and the three dynamical degrees of freedom of harmonic oscillators. In the below, however, we show that it is possible to kill the dynamical degrees of freedom of the beam by focusing on the low-frequency region effectively. 
Furthermore, we adopt the adiabatic limit for the cavity photon of the large dissipation,
$\kappa\gg\omega$, which effectively kills its dynamical degree of freedom. Then, the model will reduce to the system of the two dynamical degrees of freedom of $\delta q(\omega)$ and $\delta \Phi(\omega)$.

	The solution of Eq.~(\ref{1st_X}) under the boundary condition (\ref{bd_dxP_0}) is
	\begin{eqnarray}
		\delta X(\sigma,\omega)=A(\cos{k \sigma}-\cosh{k_e \sigma})+B(\sin{k \sigma}-\frac{k}{k_e} \sinh{k_e\sigma}),
		\label{dX_eq}
	\end{eqnarray}
	where $k$ and $k_e$ are defined as
	\begin{eqnarray}
		k&\equiv&\sqrt{\frac{-T +\sqrt{T^2+4 EI \rho \omega^2}}{2 EI}} \simeq \sqrt{\frac{\rho}{T}} \omega, \\
		k_e&\equiv&\sqrt{\frac{T +\sqrt{T^2+4 EI \rho \omega^2}}{2 EI}} 
		\simeq \sqrt{\frac{T}{EI}}.
	\end{eqnarray}
	{ The approximated formulae are valid in the 
		low-frequency region under the condition $T^2 \gg 4 |E| I \rho \omega^2$, 
		which allows us to neglect the violin modes of the beam.}
	Substituting Eq.~(\ref{dX_eq}) into the boundary conditions~(\ref{bd_dx}) and (\ref{bd_dP}), we obtain
	\begin{eqnarray}
		&&\delta X(\sigma=\ell, \omega)=A(\cos{k \ell}-\cosh{k_e \ell})+B(\sin{k \ell}-\frac{k}{k_e} \sinh{k_e \ell})=\delta x(\omega)-h \delta \Phi(\omega), 
  \\
		&&\left.\frac{\partial \delta X(\sigma, \omega)}{\partial \sigma}\right|_{\sigma=\ell}=A(-k \sin{k \ell-k_e \sinh{k_e \ell}})+B(k \cos{k \ell}-k\cosh {k_e \ell})=\delta \Phi(\omega).
	\end{eqnarray}
	These equations are written in matrix form as follows:
	\begin{eqnarray}
		\begin{pmatrix}
			\cos{k \ell}-\cosh{k_e \ell} & \sin{k \ell}-\frac{k}{k_e} \sinh k_e \ell \\
			-k \sin{k \ell}-k_e \sinh k_e \ell & k(\cos{k \ell}-\cosh{k_e \ell})\\
		\end{pmatrix}
		\begin{pmatrix}
			A\\
			B
		\end{pmatrix}
		=
		\begin{pmatrix}
			1 & -h\\
			0 & 1
		\end{pmatrix}
		\begin{pmatrix}
			\delta \x(\omega) \\
			\delta \Phi(\omega)
		\end{pmatrix} .
	\end{eqnarray} 
	The coefficients $A$ and $B$ are obtained as
	\begin{eqnarray}
		\begin{pmatrix}
			A\\
			B
		\end{pmatrix}
		=
		\begin{pmatrix}
			\barC_{00} &\barC_{01} \\
			\barC_{10} &\barC_{11}
		\end{pmatrix}
		\begin{pmatrix}
			\delta \x(\omega) \\
			\delta \Phi(\omega)
		\end{pmatrix},
	\end{eqnarray} 
	where the matrix in the right-hand side of the above equation is defined as
	\begin{eqnarray}
		\begin{pmatrix}
			\barC_{00} &\barC_{01} \\
			\barC_{10} &\barC_{11}
		\end{pmatrix}
		=
		\frac{1}{\Upsilon}
		\begin{pmatrix}
			k(\cos{k \ell}-\cosh{k_e \ell}) & -\sin{k \ell}+\frac{k}{k_e} \sinh k_e \ell -h k(\cos k \ell-\cosh k_e \ell)\\
			k \sin{k \ell}+k_e \sinh k_e \ell & \cos{k \ell}-\cosh{k_e \ell}-h(k \sin{k \ell}+k_e \sinh k_e \ell)  \\
		\end{pmatrix}
	\end{eqnarray}
	with
	\begin{eqnarray}
		\Upsilon&=&2 k(1-\cos{k \ell}\cosh {k_e \ell})+\frac{k_e^2-k^2}{k_e} \sin{k \ell} \sinh{k_e \ell}. 
	\end{eqnarray} 
	We obtain the solutions for $\delta x(\omega)$ and $\delta y(\omega)$ using Eqs.~(\ref{1st_A}) and (\ref{1st_B}) as follows:
	\begin{eqnarray}
		&&\delta \A(\omega)=\frac{2 G_0 \sqrt{n_c} \Delta}{(\kappa-i \omega)^2+\Delta^2} \delta \x(\omega)+\sqrt{2\kappa} \left\{ \frac{(\kappa-i\omega)x_{\mathrm{in}}(\omega)+\Delta y_{\mathrm{in}}(\omega)}{(\kappa-i \omega)^2+\Delta^2} \right\},
  \label{dxdqnoise}
  \\
		\quad
		&&\delta y(\omega)=\frac{1}{\Delta} \left\{(\kappa-i \omega) \delta x(\omega)-\sqrt{2\kappa} x_{\mathrm{in}}(\omega) \right\}.
	\end{eqnarray}
	From Eq.~(\ref{1st_x}), the equation of motion for the perturbation of the position of the center of mass of the mirror, $\delta q(\omega)$ is
	\begin{eqnarray}
		&&-\omega^2 \delta \x(\omega) 
		\nonumber \\
		&&~~~~~~=-\frac{T}{M} \delta \Phi(\omega)+\frac{EI}{M} \left. \frac{\partial^3 \delta X(\sigma,\omega)}{\partial \sigma^3}\right|_{\sigma=\ell}+\frac{\hbar G_0 \sqrt{n_c}}{M} \delta x(\omega) 
		\nonumber \\
		&&~~~~~~=-\frac{T}{M} \delta \Phi(\omega)+\frac{EI}{M} \{A(k^3 \sin{k \ell}-k_e^3 \sinh{k_e \ell})+B(-k^3 \cos{k \ell}-k k_e^2 \cosh{k_e \ell}) \} \nonumber \\
		&&~~~~~~+\frac{2 \hbar G^2_0 n_c \Delta}{M\{ (\kappa-i\omega)^2+\Delta^2 \}} \delta \x(\omega)+N_{\text{photon}}(\omega) 
		\nonumber \\
		&&~~~~~~=\frac{1}{M} \left[\frac{2 \hbar G^2_0 n_c \Delta}{(\kappa-i \omega)^2+\Delta^2}+EI \{\barC_{00}(k^3 \sin{k \ell}-k_e^3 \sinh{k_e \ell})- \barC_{10}(-k^3 \cos{k \ell}-k k_e^2 \cosh k_e \ell)\} \right] \delta \x(\omega) 
		\nonumber \\
		&&~~~~~~+\frac{1}{M} \left[-T +EI \{\barC_{01}(k^3 \sin{k \ell-k_e^3 \sinh{k_e \ell}})-\barC_{11}(k^3 \cos{k \ell}+k k_e^2 \cosh{k_e \ell}) \} \right] \delta \Phi(\omega)+N_{photon}(\omega) \nonumber 
		\\
		&&~~~~~~\equiv  -\omega_A^2 \delta \x(\omega) +\Delta_A^2 \delta \Phi(\omega)+N_{\text{photon}}(\omega),
	\end{eqnarray}
	where $\omega_A^2$ and $\Delta^2_A$ are defined by Eqs (\ref{Omega_A}) and (\ref{Delta_A}), and $N_{\text{photon}}(\omega)$ is defined as
	\begin{eqnarray}
		N_{\text{photon}}(\omega) \equiv \frac{\hbar G_0 \sqrt{n_c} \sqrt{2\kappa}}{M} \left\{\frac{(\kappa-i \omega)x_{\mathrm{in}}(\omega)+\Delta y_{\mathrm{in}}(\omega)}{(\kappa+i \omega)^2+\Delta^2} \right\}.
	\end{eqnarray} 
	Similarly, from Eq.~(\ref{1st_phi}), the equation of motion for the perturbation of the mirror rotation $\delta \Phi(\omega)$ can be written as follows:
	\begin{eqnarray}
		-\omega^2 \delta \Phi(\omega)&=&-\frac{h EI}{J} \left. \frac{\partial^3 \delta X(\sigma,\omega)}{\partial \sigma^3}\right|_{\sigma=\ell}+\frac{EI}{J} \left. \frac{\partial^2 \delta X(\sigma,\omega)}{\partial \sigma^2}\right|_{\sigma=\ell} \nonumber \\
		&=&-\frac{h EI}{J} \{A(k^3 \sin{k \ell}-k_e^3 \sinh{k_e \ell})+B(-k^3 \cos{k \ell}-k k_e^2 \cosh{k_e \ell}) \} \nonumber \\
		&=&-\frac{EI}{J} \{A(-k^2 \cos{k \ell}-k_e^2 \cosh{k_e \ell})+B(-k^2 \sin{k \ell}-k k_e \sinh{k_e \ell}) \} \nonumber \\
		&=&-\frac{EI}{J} \left[\barC_{00}\{h(k^3 \sin{k \ell}-k_e^3 \sinh{k_e \ell})+(-k^2 \cos{k \ell-k_e^2 \cosh{k_e \ell}})\} \right. \nonumber \\
		&&\left.+\barC_{10} \{h(-k^3 \cos{k \ell}-k k_e^2 \cosh{k_e \ell})+(-k^2 \sin{k \ell}-k k_e \sinh{k_e \ell}) \}  \right] \delta x(\omega) \nonumber \\
		&&-\frac{EI}{J} \left[ \barC_{01} \{h(k^3 \sin{k \ell}-k_e^3 \sinh{k_e \ell})+(-k^2 \cos{k \ell-k_e^2 \cosh{k_e \ell}})\} \right. \nonumber \\
		&&\left. +\barC_{11} \{h(-k^3 \cos{k \ell}-k k_e^2 \cosh{k_e \ell})+(-k^2 \sin{k \ell}-k k_e \sinh{k_e \ell}) \}  \right] \delta \Phi(\omega)  \nonumber \\
		&\equiv & \omega_B^2 \delta \x(\omega) -\Delta_B^2 \delta \Phi(\omega).
	\end{eqnarray}
	From these equations, we can obtain
	\begin{align}
		-\omega^2
		\begin{pmatrix}
			\delta \x(\omega)\\
			\delta \Phi (\omega)
		\end{pmatrix}
		=\begin{pmatrix}
			-\omega_A^2 & \Delta_A^2 \\
			\omega_B^2 & -\Delta_B^2
		\end{pmatrix}
		\begin{pmatrix}
			\delta \x(\omega) \\
			\delta \Phi(\omega)
		\end{pmatrix}
		+\begin{pmatrix}
			N_{\text{photon}}(\omega) \\
			0
		\end{pmatrix},
		\label{eom_pnoise}
	\end{align}
	where $\omega_A^2, \Delta_A^2,\omega_B^2, \Delta_B^2$ are: 
	\begin{eqnarray}
		\omega_A^2&=&-\frac{1}{M}\left[ -\frac{EI k}{\det C} (k^2+k_e^2) (k \sin{k \ell}\cosh{k_e \ell}+k_e \cos{k\ell}\sinh{k_e \ell})+\frac{2 \hbar G^2_0 n_c \Delta}{(\kappa-i \omega)^2+\Delta^2}\right], \label{Omega_A}\\
		\Delta_A^2 &=&\frac{1}{M}\biggl[ -T+
		\frac{EI}{\det C} \biggl\{ k(k_e^2-k^2) +\frac{1}{k_e}(k_e^4+k^4) \sin{k \ell}\sinh{k_e \ell}-k(k_e^2-k^2)\cos{k\ell}\cosh{k_e \ell}
  \label{Delta_A}
  \nonumber \\
		&&\hspace{95pt} +h k(k^2+k_e^2)(k \sin{k \ell}\cosh{k_e \ell}+k_e \cos{k\ell}\sinh{k_e \ell}) \biggr\} \biggr], \\ 
		\omega_B^2 &=& -\frac{EI}{J} \frac{k}{\det C} \left\{(k_e^2-k^2) -2k k_e \sin{k \ell}\sinh{k_e \ell}-(k_e^2-k^2)\cos{k\ell}\cosh{k_e \ell}\right. \nonumber \\
		&&\hspace{65pt} \left. -h (k^2+k_e^2)(k \sin{k \ell}\cosh{k_e \ell}+k_e \cos{k\ell}\sinh{k_e \ell}) \right\}, \\ 
		\Delta_B^2 &=& \frac{EI}{J}\frac{1}{\det C} \frac{k^2+k_e^2}{k_e}\left[k_e (1+h^2 k^2)\sin{k \ell}\cosh{k_e \ell} +\{k(-1+h^2 k_e^2)\cos{k\ell}+h(k^2+k_e^2)\sin{k \ell} \} \sinh{k_e \ell} \right]. 
		\label{Delta_B}
	\end{eqnarray}
We note that $\omega_A^2$, $\Delta_A^2$, $\omega_B^2$, and $\Delta_B^2$ depend on $\omega$. This is because the mirror is coupled to the beam described by $\delta X(\sigma,t)$, which has infinite numbers of degrees of freedom. Therefore, it is not trivial whether the mirror's modes are described by the variables of the mirror $\delta q$ and $\delta \Phi$. 
However, we are interested in the low-frequency region of the resonant frequency of the mirror, which is found under the condition $\omega^2 \ll T^2/4|E|I \rho$. This condition reduces Eqs.~(\ref{Omega_A})$\sim$(\ref{Delta_B}) to
		\begin{eqnarray}
			&&\omega_A^2 \simeq -\frac{1}{M} \left\{ \frac{2 \hbar G^2_0 n_c \Delta}{\kappa^2+\Delta^2}-\frac{T}{\ell}\left(1+\frac{2}{\ell} \sqrt{\frac{EI}{T}} \right) \right\}, 
			\label{Omega_Ap} \\
			&&\Delta_A^2 \simeq \frac{Th}{M \ell} \left\{1+\left(\frac{1}{h}+\frac{2}{\ell}\right)  \sqrt{\frac{EI}{T}}  \right\}, 
			\label{Delta_Ap}\\
			&&\omega_B^2 \simeq \frac{Th}{J \ell} \left\{1+\left(\frac{1}{h}+\frac{2}{\ell}\right) \sqrt{\frac{EI}{T}}  \right\}, 
			\label{Omega_Bp} \\
			&&\Delta_B^2 \simeq  \frac{Th}{J} \left\{1+\frac{h}{\ell}+\left(\frac{1}{h}+\frac{2}{\ell}\right) \sqrt{\frac{EI}{T}}  \right\}. 
			\label{Delta_Bp} 
		\end{eqnarray}
Thus, in the low-frequency region $\omega^2 \ll T^2/4|E|I \rho$, Eqs.~(\ref{Omega_Ap})$\sim$(\ref{Delta_Bp}) are independent of $\omega$, which implies that the theory can be reduced to a two-mode theory of $\delta \x$ and $\delta \Phi$. 
In the two-mode theory, the violin modes of the beam are neglected, but the 
internal friction of the beam can be included in the complex Young's modulus. 
The complex Young's modulus makes small imaginary parts in the quantities, $\omega_A^2$, $\Delta_A^2$, $\omega_B^2$, and $\Delta_B^2$. In addition, we also assumed that the cavity photons dissipated faster than the oscillations of the mirror $\kappa \gg \omega$, in our analysis.

\subsection{Frequency {dependence} of the dissipation \label{beamD}}
From Eq.~(\ref{eom_pnoise}), by eliminating the variable $\delta\Phi(\omega)$, we obtain the following result with respect to $\delta \x(\omega)$:
	\begin{eqnarray}
		\delta \x(\omega)
		=
		\frac{\omega^2 -\Delta_B^2}{f(\omega)} N_{\text{photon}}(\omega),
	\end{eqnarray}
	where we defined
	\begin{eqnarray}
		f(\omega)\equiv
		{-(\omega^2-\omega_A^2)(\omega^2-\Delta_B^2)+\omega_B^2 \Delta_A^2}.
	\end{eqnarray} 
We now focus on the solutions that satisfy $f(\omega)=0$ because they 
provide the normal modes of the coupled oscillators and determine the resonant
oscillations against fluctuating noise forces, which must be related to 
dissipation through the fluctuation-dissipation relation.
Our model enables us to investigate the effects of the degrees of freedom 
of the beam on the motion of the mirror in the low-frequency region. 
In particular, the dissipation owing to the internal friction 
of the beam can be investigated using the complex Young's modulus 
$E(=E_0(1- i\phi))$.  
Subsequently, we focus on the dissipation of mirror motion 
based on the complex Young's modulus. 
	
The solutions of the equation $f(\omega)=0$ are complex owing to the presence of dissipation.
The real parts of the solutions correspond to the eigenfrequencies of the normal modes
of the coupled oscillators. 
The imaginary parts of the solutions represent the dissipation in each mode.
Here, we assume that the internal loss factor $\phi$ is very small, namely, $\phi \ll 1$, and the imaginary parts of the solutions are proportional to the first order of $\phi$.
To obtain the solution, we rewrite the function $f(\omega)$ as
	\begin{eqnarray}
		f(\omega)
		\approx
		-(\omega^2-\omega_{AR}^2 -i \phi \omega_{AI}^2)(\omega^2-\Delta_{BR}^2-i \phi \Delta_{BI}^2)+(\Delta_{AR}^2+i \phi \Delta_{AI}^2) (\omega_{BR}^2+i \phi \omega_{BI}^2),
	\end{eqnarray} 
	where we assume that 
	$
	\omega_{A}^2=\omega_{AR}^2 +i \phi \omega_{AI}^2, 
	$
	$
	\omega_{B}^2=\omega_{BR}^2 +i \phi \omega_{BI}^2,
	$ 
	$\Delta_{A}^2=\Delta_{AR}^2 +i \phi \Delta_{AI}^2,
	$ 
	$\Delta_{B}^2=\Delta_{BR}^2 +i \phi \Delta_{BI}^2,
	$
	and explicitly define by assuming $\phi\ll 1$
	\begin{align}
		\omega_{AR}^2
		&=
		-\frac{1}{M}\left\{-\frac{T}{\ell}\left(1+\frac{2}{\ell}\sqrt{\frac{E_0 I}{T}} \right)+\frac{\hbar \Delta G^2}{\kappa^2+\Delta^2} \right\}, 
		\quad
		\omega_{AI}^2
		=
		-\frac{T}{M \ell^2}\sqrt{\frac{E_0 I}{T}}, 
		\\ 
		\Delta_{AR}^2
		&=
		\frac{T h}{M \ell} \left\{1+ \left(\frac{1}{h}+\frac{2}{\ell}\right) \sqrt{\frac{E_0 I}{T}}\right\},
		\quad
		\Delta_{AI}^2
		=-\frac{(2h+\ell)T}{2 M \ell^2}\sqrt{\frac{E_0 I}{T}}, 
		 \\ 
		\omega_{BR}^2
		&=
		\frac{T h}{J \ell} \left\{1+\left(\frac{1}{h}+\frac{2}{\ell} \right)\sqrt{\frac{E_0 I}{T}} \right\},
		\quad
		\omega_{BI}^2
		=
		-\frac{(2h+\ell)T}{2 J \ell^2}\sqrt{\frac{E_0 I}{T}}, 
		\\ 
		\Delta_{BR}^2
		&=
		\frac{T h}{J} \left\{1+\frac{h}{\ell}+\left(\frac{1}{h}+\frac{2}{\ell}\right) \sqrt{\frac{E_0 I}{T}} \right\}, 
		\quad
		\Delta_{BI}^2=-\frac{(2h +\ell)T}{2 J \ell}\sqrt{\frac{E_0I}{T}}. 
	\end{align}
	Assuming that the solution of the equation $f(\omega_{\text{c}})=0$ is $\omega_c=\omega_0+i \phi \omega_1$, $f(\omega_c)$ can be expressed as: 
	\begin{eqnarray}
		&&f(\omega_c)
		=-(\omega_0^2- \omega_{AR}^2)(\omega_0^2-\Delta_{BR}^2)+\Delta_{AR}^2 \omega_{BR}^2 
		+i \phi \Bigl\{(\omega_0^2-\Delta_{BR}^2)(-2 \omega_0 \omega_1 +\omega_{AI}^2)
		\nonumber
		\\
		&&~~~~~~~~~~
		+(\omega_0^2-\omega_{AR}^2)(-2 \omega_0  \omega_1+\Delta_{BI}^2)+\Delta_{AR}^2 \omega_{BI}^2 +\Delta_{AI}^2 \omega_{BR}^2 \Bigr\}.
	\end{eqnarray}
	Comparing the real and imaginary parts of the equation $f(\omega_c)=0$, we obtain the following conditions:
	\begin{eqnarray}
		&&(\omega_0^2- \omega_{AR}^2)(\omega_0^2-\Delta_{BR}^2)-\Delta_{AR}^2 \omega_{BR}^2=0,
		\\
		&&(\omega_0^2-\Delta_{BR}^2)(-2 \omega_0 \omega_1 +\omega_{AI}^2)+(\omega_0^2-\omega_{AR}^2)(-2 \omega_0  \omega_1+\Delta_{BI}^2)+\Delta_{AR}^2 \omega_{BI}^2 +\Delta_{AI}^2 \omega_{BR}^2 =0.
	\end{eqnarray}
	{
		Then, $\omega_{0}$ and $\omega_{1}$ are obtained as follows: 
		\begin{align}
			\omega_{0}^2
			&=
			\frac{(\Delta_{BR}^2+\omega_{AR}^2)\pm \sqrt{(\omega_{AR}^2-\Delta_{BR}^2)^2+4\Delta_{AR}^2 \omega_{BR}^2}}{2} 
			\equiv \omega_{0\pm}^2,
			\\
			\omega_{1}
			&=
			\frac{-\omega_{0}^2(\Delta_{BI}^2+\omega_{AI}^2)+\Delta_{BR}^2 \omega_{AI}^2+\Delta_{BI}^2 \omega_{AR}^2-\Delta_{AR}^2 \omega_{BI}^2-\Delta_{AI}^2 \omega_{BR}^2 }{2\omega_{0}(2\omega_{0}^2-\omega_{AR}^2-\Delta_{BR}^2)}.
			\label{omega_1}
	\end{align} }

The resonant frequency $\omega^2_{0}$ has two modes, $\omega^2_{0+}$ and $\omega^2_{0-}$.
		\begin{figure}[t]
			\centering
			\includegraphics[width=100mm]{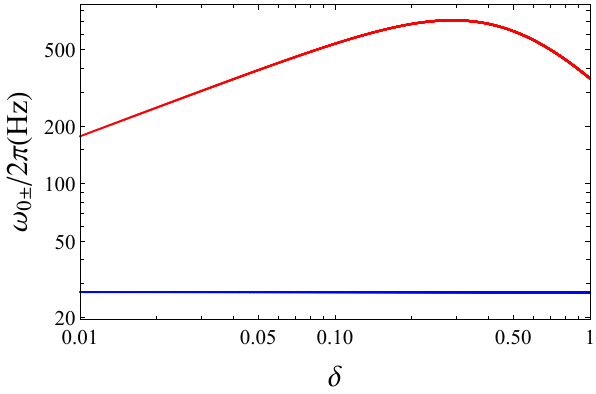}
			\caption{
Resonant frequency behavior as a function of the  normalized detuning $\delta(=-\Delta/2\kappa)$.
The red curve and blue line represent the resonance frequency of the translational (pendulum) mode $\omega_{0+}$ and the resonance frequency of the rotational mode $\omega_{0-}$, respectively.
				\label{fig:freq}}
		\end{figure}
For $(\Delta^2_{BR}-\omega^2_{AR})^2 \gg \Delta_{AR}^2 \omega_{BR}^2$,  the resonant frequencies $\omega^2_{0\pm}$ are approximated as
		\begin{eqnarray}
			&&\omega^2_{0+}
			\approx
			\frac{(\Delta_{BR}^2+\omega_{AR}^2)+ |(\omega_{AR}^2-\Delta_{BR}^2)|}{2}
			= \omega_{AR}^2,
			\\
			&&\omega^2_{0-}
			\approx
			\frac{(\Delta_{BR}^2+\omega_{AR}^2)- |(\omega_{AR}^2-\Delta_{BR}^2)|}{2} 
			= \Delta_{BR}^2.
		\end{eqnarray} 
In this study, we focus on the case $(\Delta^2_{BR}-\omega^2_{AR})^2 \gg \Delta_{AR}^2 \omega_{BR}^2$. In this case, $\delta \x$ and $\delta \Phi$ correspond to the translational pendulum motion of the center of mass and the rotational motion of the mirror around the center of mass, which allows us to refer to $\delta \x$ and $\delta \Phi$ as the pendulum mode and the  rotational modes, respectively.
Then, $\omega_{AR}$ and $\Delta_{BR}$ represent the resonance frequencies of the pendulum and rotational modes, respectively, although the correspondence between the two modes changes depending on the coupling. For the parameters in Table \ref{tab:parameter}, $\omega_{0+}$ and $\omega_{0-}$ correspond to the frequencies of the pendulum  and the rotational modes, respectively (see also Ref. ~\cite{Sugiyama}).
Figure~\ref{fig:freq} depicts the behavior of the resonant frequency $\omega_{0\pm}$ as a function of the normalized detuning $\delta(=-\Delta/2\kappa)$. The red and blue curves represent the pendulum and rotational modes, respectively.
		
\begin{figure}[t]
	\centering
	\includegraphics[width=100mm]{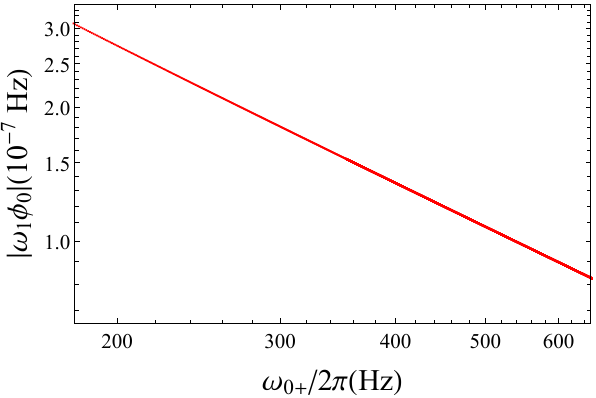}
	\caption{
Frequency response of the dissipation $\omega_{1} \phi_{0}$ at the resonant frequency of the pendulum mode of the mirror. 
		}
		\label{fig:freq_cond}
	\end{figure}

In particular, we focused on the frequency region {$180 ~\mathrm{Hz}<\omega_{0+}/2\pi<650~\mathrm{Hz}$ }, where $\omega_{0+}$ approximately corresponds to the frequency of the pendulum mode.
The frequency dependence of dissipation, expressed as $\omega_{1} \phi$, is shown in Fig.~\ref{fig:freq_cond}, where $\phi=\phi_0=10^{-3}$.
Fig.~\ref{fig:freq_cond} clearly shows that the dissipation $\omega_{1} \phi_0$ is proportional to the inverse of the frequency $1/\omega$ in the resonant frequency region of the pendulum mode.
To understand this behavior quantitatively, we can approximate the following: $\omega_{0}\approx \omega_{AR}$, which allows us to express $\omega_1$ as:
\begin{eqnarray}
		\omega_{1}
		\approx
		\frac{(-\omega_{AR}^2+\Delta_{BR}^2) \omega_{AI}^2-\Delta_{AR}^2 \omega_{BI}^2 -\Delta_{AI}^2 \omega_{BR}^2}{2 \omega_0 (\omega_{AR}^2-\Delta_{BR}^2)}
		=
		-\frac{\omega^2_{\text{AI}}}{2\omega_{0}}
		\left[
		1+\frac{\Delta_{AR}^2 \omega_{BI}^2 +\Delta_{AI}^2 \omega_{BR}^2}{(\omega_{AR}^2-\Delta_{BR}^2) \omega_{AI}^2}
		\right].
		\label{omegaonea}
	\end{eqnarray}
	The second term in the brackets on the right-hand side of Eq.~(\ref{omegaonea}) can be evaluated as
	\begin{align}
		\frac{\Delta_{AR}^2 \omega_{BI}^2 +\Delta_{AI}^2 \omega_{BR}^2}{(\omega_{AR}^2-\Delta_{BR}^2) \omega_{AI}^2}
		&=
		\frac{\Delta_{\text{AR}}\omega_{\text{BR}}}{(\omega_{AR}^2-\Delta_{BR}^2) \omega_{AI}^2}
		\frac{\Delta_{AR}^2 \omega_{BI}^2 +\Delta_{AI}^2 \omega_{BR}^2}{\Delta_{\text{AR}}\omega_{\text{BR}}}
		\nonumber\\
		\quad
		&=
		\frac{\Delta_{\text{AR}}\omega_{\text{BR}}}{(\omega_{AR}^2-\Delta_{BR}^2)}
		\Big(
		\frac{\Delta^2_{\text{AI}}}{\omega^2_{\text{AI}}}
		\Big)
		\Big(
		\frac{\omega_{\text{BR}}}{\Delta_{\text{AR}}}
		\Big)
		\Big(
		1+\frac{\omega^2_{\text{BI}}}{\Delta^2_{\text{AI}}}\frac{\Delta^2_{\text{AR}}}{\omega^2_{\text{BR}}}
		\Big)
		\nonumber\\
		\quad
		&=
		\ell\sqrt{\frac{M}{J}}\Big(1+\frac{2h}{\ell}\Big)\frac{\Delta_{\text{AR}}\omega_{\text{BR}}}{(\omega_{AR}^2-\Delta_{BR}^2)}.
		\label{secomega1}
\end{align}
Note that the condition $\omega_{0} \approx \omega_{\text{AR}}$ is justified when the condition $(\omega_{AR}^2-\Delta_{BR}^2)^2 \gg \Delta^2_{\text{AR}}\omega^2_{\text{BR}}$ holds.
Thus, the condition $(\Delta_{AR}^2 \omega_{BI}^2 +\Delta_{AI}^2 \omega_{BR}^2)/(\omega_{AR}^2-\Delta_{BR}^2) \omega_{AI}^2 \ll 1$ is valid as long as $\ell \sqrt{M/J}(1+2h/\ell) \sim \mathcal{O}(1)$, which is satisfied by our parameters.

We define the mechanical decay rate of the pendulum mode by 
\begin{eqnarray}
\Gamma_{\text{r}}(\omega)=\phi_0 \omega_1.
\end{eqnarray}
As is shown in Fig.~\ref{fig:freq_cond}, $\Gamma_{\text{r}}(\omega)\propto 1/\omega$. Then, we can write the relation
	\begin{eqnarray}
		\Gamma_{\text{r}} (\omega)\approx {\Gamma_{\text{r}}(\Omega) \Omega\over\omega},
		\label{sd}
	\end{eqnarray}
where $\Omega$ is the frequency of the pendulum mode in the absence of the cavity light. 
Thus, the relation Eq.(\ref{sd}),  which is called the structural damping relation
can be explained by the internal friction in the beam mode modeled as a complex Young's modulus $\phi$.
In the present paper, we use the value of $\Gamma_{\text{r}}(\Omega)$ in Table \ref{tab:parameter}, which is obtained by $\Gamma(\Omega)$ multiplied by $4.18$, 
where $\Gamma(\Omega)$ is the mechanical decay rate in the absence of 
rotational mode \cite{Matsumoto} in Table \ref{tab:parameter}. 
The factor of $4.18$ is obtained from
$\Gamma_r(\Omega)\approx\Gamma_r(\omega)\omega/\Omega\approx\phi_0\omega_{AI}^2/\Omega$
and
the value of $\Gamma(\Omega)$ in Table \ref{tab:parameter}. 
This factor is consistent with the result in Ref.~\cite{Sugiyama}, 
which demonstrated that the quality factor is reduced by the 
factor $4$ due to the presence of the internal friction of the beam
through the rotational mode.

\subsection{Noise term \label{beamE}}
In this subsection, we consider the equations of motion for the two-mode theory including the noise. 
Because we assume a very low-pressure $10^{-5} \mathrm{~Pa}$ environment for our system, the residual gas friction can be neglected and the structural damping model owing to the internal structure of the beam can be dominantly employed \cite{Matsumoto,Sugiyama}.
Therefore, we add structural damping and random thermal noise that satisfy the fluctuation-dissipation relation in the equation of motion for $\delta \x(\omega)$.
We also considered adding feedback cooling, which effectively decreases the friction force, by detecting the light exiting the cavity.
In this case, Eq.~(\ref{1st_q}) is replaced with the effective equation
	\begin{eqnarray}
		-M \omega^2 \delta \x (\omega)
		=
		-T \delta\Phi(\omega) +E_0 I\left. \frac{\partial^3 \delta X(\omega)}{\partial \sigma^3}\right|_{\sigma=\ell}+\hbar G_0 \sqrt{n_c} \delta x(\omega) +i \omega \Gamma_{\text{r}} M \delta \x (\omega)+\xi(\omega)-g_{\mathrm{FB}}(\omega) X_A(\omega), \label{noiseterm_x}
\end{eqnarray}
where $\Gamma_{\text{r}}$ is the mechanical (structural) decay rate defined at the end of the previous subsection, which describes the effect of the internal friction of the beam through the rotational mode.
The last term in Eq.~(\ref{noiseterm_x}) represents feedback cooling and $X_{\text{A}}$ is the measured output light, which is defined as
	\begin{eqnarray}
		X_{\text{A}}(\omega)=\sqrt{\eta} \A_{\mathrm{out}}(\omega)+\sqrt{1-\eta}\A_{\mathrm{in}}'(\omega),
		\quad
		\A_{\mathrm{out}}(\omega)
		=
		\A_{\mathrm{in}}(\omega)-\sqrt{2 \kappa} \delta x(\omega), 
	\end{eqnarray}
	where $\eta \in [0,1]$ represents the detection efficiency and $x_{\mathrm{in}}'(\omega)$ is a false vacuum, indicating imperfect detection.
	In addition, $\xi$ is the fluctuation force that satisfies the time average of $\langle \xi(t) \rangle=0$.
	The correlation function with respect to $\xi$ is extended to a finite dimension in Ref.~\cite{Genes,Giovannetti},
	\begin{eqnarray}
		\langle \xi(t) \xi(t') \rangle=\int \frac{d \omega}{2\pi} e^{-i \omega(t-t')}  M \hbar \Gamma_{\text{r}} \omega \left[\coth \left(\frac{\hbar \omega}{2 k_B T_0}\right)+1 \right],
		\label{cor0}
	\end{eqnarray}
	or
	\begin{eqnarray}
		\langle \xi(\omega) \xi(\omega') \rangle= M \hbar \Gamma_{\text{r}} \omega \left[\coth \left(\frac{\hbar \omega}{2 k_B T_0}\right)+1 \right]2\pi\delta(\omega-\omega').
		\label{cor1}
	\end{eqnarray}
Here $T_0$ is the ambient temperature.  
 We assume that 
the random thermal noise for the perturbation of the mirror rotation $\delta \Phi(\omega)$ is supposed to be small and we neglected it. We adopt this assumption in the present paper because we focus our research on the dissipation in the pendulum mode from the beam's internal friction.

Eq.~(\ref{noiseterm_x}) with the feedback cooling can be treated by \begin{eqnarray}
-M \omega^2 \delta \x (\omega)
=
-T \delta\Phi(\omega) +E_0 I\left. \frac{\partial^3 \delta X(\omega)}{\partial \sigma^3}\right|_{\sigma=\ell}+\hbar G_0 \sqrt{n_c} \delta x(\omega) +i \omega \gamma_m M \delta \x (\omega)+\xi_m(\omega),
\end{eqnarray}
where $\gamma_m$ is the effective damping rate of the mechanical oscillator, 
corresponding to the effective temperature at the center of mass of the mirror, decreasing to $T_0 \Gamma_{\text{r}}/\gamma_m$.
The fluctuation force $\xi$ also changes owing to feedback cooling and is redefined as $\xi_{\text{m}}$, whose correlation functions~\eqref{cor0} and \eqref{cor1} are
	\begin{eqnarray}
	\langle \xi_m(t) \xi_m(t') \rangle= \int \frac{d \omega}{2\pi} e^{-i \omega(t-t')} M \hbar \gamma_m \omega \left[\coth \left(\frac{\hbar\omega{\gamma_m}}{2 k_B T_0{\Gamma_{\text{r}}}}\right)+1 \right],
	\label{xi_m^2_avg}
\end{eqnarray} 
or
\begin{eqnarray}
\langle \xi_m(\omega) \xi_m(\omega') \rangle=  M \hbar \gamma_m \omega \left[\coth \left(\frac{\hbar\omega{\gamma_m}}{2 k_B T_0{\Gamma_{\text{r}}}}\right)+1 \right]
2\pi\delta(\omega-\omega').
\label{xi_m^2_avgb}
\end{eqnarray} 

Including the feedback cooling and noise terms, Eq.~\eqref{eom_pnoise} leads to:
\begin{align}
			-\omega^2
			\begin{pmatrix}
				\delta \x(\omega)\\
				\delta \Phi (\omega)
			\end{pmatrix}
			&=
			\begin{pmatrix}
				-\omega^2_\text{AR}+ i \gamma_m \omega & \Delta^2_{\text{AR}} \\
				\omega^2_\text{BR} & -\Delta^2_\text{BR}
			\end{pmatrix}
			\begin{pmatrix}
				\delta \x(\omega) \\
				\delta \Phi(\omega)
			\end{pmatrix}
			+
			\frac{1}{M}\begin{pmatrix}
				\frac{\hbar G_0 \sqrt{n_c} \sqrt{2 \kappa}}{(\kappa^2+\Delta^2)} (\kappa \A_{\mathrm{in}}(\omega)+\Delta \B_{\mathrm{in}}(\omega))+{\xi_m(\omega)}
				\\
				0
			\end{pmatrix},
			\label{1st_dis_phi}
\end{align}
which yields the following solutions
\begin{eqnarray}
\delta \x(\omega)
			&=&
			\frac{\omega^2
				-\Delta_{BR}^2}{F(\omega)} \left\{
			\frac{\hbar G_0 \sqrt{n_c} \sqrt{2 \kappa}}{M(\kappa^2+\Delta^2)} (\kappa \A_{\mathrm{in}}(\omega)+\Delta \B_{\mathrm{in}}(\omega))+\frac{\xi_m(\omega)}{M} \right\},
			\label{sol_dx_o}
			\\
			\delta \Phi(\omega)
			&=&
			-\frac{\omega_{BR}^2 }{\omega^2-\Delta_{BR}^2}\delta \x(\omega)
			\nonumber\\
			&=&
			-\frac{\omega_{BR}^2}{F(\omega)} \left\{\frac{\hbar G_0 \sqrt{n_c} \sqrt{2 \kappa}}{M(\kappa^2+\Delta^2)} (\kappa \A_{\mathrm{in}}(\omega)+\Delta \B_{\mathrm{in}}(\omega))+\frac{\xi_m(\omega)}{M} \right\},
\label{sol_dPhi_o}
\end{eqnarray}
where $F(\omega)$ is defined as
\begin{eqnarray}
F(\omega)=-(\omega^2+i \gamma_m \omega -\omega_{AR}^2)(\omega^2
-\Delta_{BR}^2)+\Delta_{AR}^2 \omega_{BR}^2.
\end{eqnarray}
Finally, using the above results, we summarize the solutions with the noise terms for $\delta p(\omega)$, 
$\delta \Pi_{\Phi}(\omega)$, and $X_{\text{A}}(\omega)$ as
\begin{align}
&\delta p(\omega)
		=
		-i M \omega \delta \x(\omega), 
\label{sol_dp_o}
		\\
&\delta \Pi_{\Phi}(\omega)
		=
		-i J \omega \delta \Phi(\omega),
\label{sol_dPi_o}
		\\
&X_A(\omega)
=
\sqrt{\eta} \left[ \left\{1-\frac{2 \kappa^2}{\kappa^2+\Delta^2} -\frac{\omega^2
-\Delta_{BR}^2}{F(\omega)} \frac{4 \hbar G_0^2 n_c \Delta \kappa^2}{M(\kappa^2+\Delta^2)^2} \right\}\A_{\mathrm{in}}(\omega)
\right. 
-\left\{\frac{2 \kappa \Delta}{\kappa^2+\Delta^2}+\frac{\omega^2
-\Delta_{BR}^2}{F(\omega)} \frac{4 \hbar G_0^2 n_c \Delta^2 \kappa}{M(\kappa^2+\Delta^2)^2} \right\} \B_{\mathrm{in}}(\omega) 
\nonumber \\
&~~~~~~~~~
\left. -\frac{2 G_0 \sqrt{n_c} \Delta \sqrt{2 \kappa}}{\kappa^2+\Delta^2} \frac{\omega^2
-\Delta_{BR}^2}{F(\omega)} \frac{\xi_m(\omega)}{M}
\right] 
+\sqrt{1- \eta}\A_{\mathrm{in}}'(\omega). \label{XA_sol_noise}
\end{align}
Here, $\delta q(\omega)$ and $\delta\Phi(\omega)$ in Eqs.~(\ref{sol_dp_o}) and (\ref{sol_dPi_o}) are given by Eqs.~(\ref{sol_dx_o}) and (\ref{sol_dPhi_o}), respectively.

\section{Wiener filter and conditional covariance\label{wiener}}
In this section, we introduce an optimal filter, the Wiener filter, that minimizes the root-mean-square error between the true and estimated values during the process of obtaining an estimate from the observed physical quantities \cite{Wiener,NYamamoto}. We investigated the conditional variance of the mirror's motion under continuous measurement and feedback control based on the two-mode theory developed in the previous section, which is a generalization of the works in \cite{MY,Meng} to the multi-mode theory. 
One can see a reference, e.g., Ref.~\cite{Wiener,NYamamoto,Grover}, for a nice review and textbook of the Wiener filter. 
In the first part of the section, we present the derivation of the Wiener filtering process to minimize the uncertainty and the conditional covariance in a self-contained way.

\subsection{Spectral density and Wiener filter}
The Wiener filter is an optimal filter that minimizes the mean-square difference between the true and estimated values when obtaining an estimate from an observed physical quantity. 
We consider the process of estimating a physical quantity of interest from an observed physical quantity in Fourier space. 
Because a perfect observation is impossible, the true value of the physical quantity and that estimated from the observed value will always be different. 
Therefore, we applied a filter that minimized the mean square difference between the true and estimated values. 
Let $\delta q (\omega)$ be the true value, $X_A (\omega)$ be the observed quantity, $H (\omega)$ be the filter that allows optimal estimation, and $\delta q_e (\omega)$ be the estimated value from the observed value. $\delta q_e(\omega)$ is expressed as
	\begin{eqnarray}
		\delta q_{\text{e}}(\omega)=H(\omega) X_A(\omega).
  \label{dqe}
	\end{eqnarray}
We define the mean-square difference between the true $\delta q(\omega)$ and estimated $\delta q_{\text{e}}(\omega)$ values in Fourier space as\footnote{
Assuming $A(t)$ and $B(t)$ are Hermite operators, we may write
\begin{eqnarray}
 \langle A(t)B(t)+B(t)A(t)\rangle
 &\equiv&\langle A(t)^\dagger B(t)\rangle
 +\langle B(t) A(t)^\dagger\rangle
 \nonumber\\
 &=&{1\over (2\pi)^2}\int\int d\omega d\omega'
 (\langle A(\omega)^\dagger B(\omega')\rangle e^{i(\omega -\omega')t}
 +
 \langle B(\omega) A(\omega')^\dagger\rangle e^{-i(\omega -\omega')t})
\nonumber \\
&=&{1\over (2\pi)^2}\int\int d\omega d\omega'
 (\langle A(\omega)^\dagger B(\omega')\rangle 
 +
 \langle B(\omega') A(\omega)^\dagger \rangle)e^{i(\omega -\omega')t}.
 \nonumber
\end{eqnarray}
When we introduce the spectral density $S_{AB}(\omega)$ by
$\langle A^{\dag}(\omega)B(\omega')+B(\omega')A^{\dag}(\omega) \rangle
=2\pi \delta(\omega-\omega')S_{AB}(\omega)$,
we have
\begin{eqnarray}
 \langle A(t)B(t)+B(t)A(t)\rangle
={1\over2\pi }\int d\omega S_{AB}(\omega).
\nonumber
\end{eqnarray}
Therefore, (twice of) the square of an operator $A(t)$ in Fourier space 
can be defined as 
\begin{eqnarray}
 \langle A(\omega)^\dagger A(\omega)+A(\omega)A(\omega)^\dagger\rangle.
 \nonumber
\end{eqnarray}
}

\begin{eqnarray}
		\langle|\delta q(\omega)-\delta q_{\text{e}}(\omega)|^2\rangle
		&\equiv&\langle (\delta q(\omega)-H(\omega) X_A(\omega))^{\dag} (\delta q(\omega)-H(\omega) X_A(\omega)) \nonumber \\
		&&+(\delta q(\omega)-H(\omega) X_A(\omega))(\delta q(\omega)-H(\omega) X_A(\omega))^{\dag}\rangle \nonumber \\
		&=&\langle \delta q^{\dag}(\omega) \delta q(\omega)+\delta q(\omega) \delta q^{\dag}(\omega)\rangle -H^*(\omega) \langle X_A^{\dag}(\omega) \delta q(\omega)+\delta q(\omega) X_A^{\dag}(\omega)\rangle \nonumber \\
		&&-H(\omega) \langle \delta q^{\dag}(\omega) X_A(\omega)+ X_A(\omega) \delta q^{\dag}(\omega)\rangle \nonumber \\
		&&+H(\omega) H^*(\omega) \langle X_A^{\dag}(\omega) X_A(\omega)+X_A (\omega) X_A^{\dag}(\omega)\rangle.
		\label{detect_1}
\end{eqnarray}
We find the minimum of $\langle|\delta q(\omega)-\delta q_{\text{e}}(\omega)|^2\rangle$. To this end, its partial derivative with respect to the filter $H(\omega)$ is
\begin{eqnarray}
		\frac{\partial \langle|\delta q(\omega)-\delta q_{\text{e}}(\omega)|^2\rangle}{\partial H(\omega)} =- \langle \delta q^{\dag}(\omega) X_A(\omega)+ X_A(\omega) \delta q^{\dag}(\omega)\rangle +H^*(\omega) \langle X_A^{\dag}(\omega) X_A(\omega)+X_A(\omega) X_A^{\dag}(\omega)\rangle. 
\end{eqnarray}
Here we regarded that the filter $H(\omega)$ is independent of its complex conjugate $H^*(\omega)$ because it has two degrees of freedom. This technique is used to derive the equation of motion of a complex scalar field with the variational principle from the action integral.
Here, the filter $H_{\text{opt}}(\omega)$ that performs the best estimation minimizes the difference between the true and estimated values $\langle | \delta q(\omega)-\delta q_{\text{e}}(\omega)|^2\rangle$. Thus
\begin{eqnarray}
		\frac{\partial \langle|\delta q(\omega)-\delta q_{\text{e}}(\omega)|^2\rangle}{\partial H_{\text{opt}}(\omega)}=0
\end{eqnarray}
is satisfied. The complex conjugate $H_{\text{opt}}^*(\omega)$ of the filter for the optimal estimation is given by
\begin{eqnarray}
		H_{\text{opt}}^*(\omega)=\frac{\langle \delta q^{\dag}(\omega) X_A(\omega)+ X_A(\omega) \delta q^{\dag}(\omega)\rangle}{\langle X_A^{\dag}(\omega) X_A(\omega)+X_A(\omega) X_A^{\dag}(\omega)\rangle}.
\end{eqnarray}
In addition, when choosing $H(\omega)$ as the filter $H_{\rm opt}(\omega)$ that facilitates optimal estimation, substituting into equation (\ref{detect_1}), we obtain
\begin{eqnarray}
		\langle |\delta q(\omega)-\delta q_{\text{e}}(\omega)|^2 \rangle
		&=&\langle \delta q^{\dag}(\omega) \delta q(\omega)+\delta q(\omega) \delta q^{\dag}(\omega)\rangle \nonumber \\
		&&-\frac{\langle \delta q^{\dag}(\omega) X_A(\omega)+ X_A(\omega) \delta q^{\dag}(\omega)\rangle}{\langle X_A^{\dag}(\omega) X_A(\omega)+X_A(\omega) X_A^{\dag}(\omega)\rangle} \langle X_A^{\dag}(\omega) \delta q(\omega)+\delta q(\omega) X_A^{\dag}(\omega)\rangle \nonumber \\
		&&-H_{\text{opt}}(\omega) \langle \delta q^{\dag}(\omega) X_A(\omega)+ X_A(\omega) \delta q^{\dag}(\omega)\rangle \nonumber \\
		&&+H_{\text{opt}}(\omega) \frac{\langle \delta q^{\dag}(\omega) X_A(\omega)+ X_A(\omega) \delta q^{\dag}(\omega)\rangle}{\langle X_A^{\dag}(\omega) X_A(\omega)+X_A(\omega) X_A^{\dag}(\omega)\rangle} \langle X_A^{\dag}(\omega) X_A(\omega)+X_A(\omega) X_A^{\dag}(\omega)\rangle \nonumber \\
		&=&\langle \delta q^{\dag}(\omega) \delta q(\omega)+\delta q (\omega) \delta q^{\dag}(\omega) \rangle -|H_{opt}(\omega)|^2 \langle X_A^{\dag}(\omega) X_A(\omega)+ X_A(\omega) X_A^{\dag}(\omega)\rangle. 
\end{eqnarray}
Here we use
\begin{eqnarray}
	H_{\text{opt}}(\omega)=\frac{\langle X_A^{\dag}(\omega) \delta q(\omega)+\delta q(\omega) X_A^{\dag}(\omega)\rangle}{\langle X_A^{\dag}(\omega)X_A(\omega)+X_A(\omega) X_A^{\dag}(\omega)\rangle}.
	\end{eqnarray}
In general, the symmetrized one sided spectral density $S_{AB}(\omega)$ for some physical quantity $A(\omega),B(\omega)$ is defined as $2\pi \delta(\omega-\omega')S_{AB}(\omega)\equiv\langle A^{\dag}(\omega)B(\omega')+B(\omega')A^{\dag}(\omega) \rangle$\cite{Miao10}.
Using this equation, it is represented as
\begin{eqnarray}
H_{\text{opt}}(\omega)=\frac{S_{X_A \delta q}(\omega)}{S_{X_A X_A}(\omega)}.
\end{eqnarray}
The filter $H_{\delta q}(\omega)$ which minimizes the mean square difference between the true value $\delta q$ and the estimated value $\delta q_{\text{e}}(\omega)$ is called the Wiener filter.
We can determine the Wiener filters for the other variables in a similar manner. 
		
To determine the Wiener filter, we must consider the causal part of $H_{\rm opt}(\omega)$, 
\begin{align}
H_{\delta \x}(\omega)
&=
\frac{1}{S_{X_A X_A}^+(\omega)} \left[\frac{S_{X_A \delta \x}(\omega)}{S_{X_A X_A}^{-}(\omega)} \right]_{+},
\quad
H_{\delta p}(\omega)=\frac{1}{S_{X_A X_A}^+(\omega)} \left[\frac{S_{X_A \delta p}(\omega)}{S_{X_A X_A}^{-}(\omega)} \right]_{+},
 \\
H_{\delta \Phi}(\omega)
&=
\frac{1}{S_{X_A X_A}^+(\omega)} \left[\frac{S_{X_A \delta \Phi}(\omega)}{S_{X_A X_A}^{-}(\omega)} \right]_{+},
\quad
H_{\delta \Pi_{\Phi}}(\omega)=\frac{1}{S_{X_A X_A}^+(\omega)} \left[\frac{S_{X_A \delta \Pi_{\Phi}}(\omega)}{S_{X_A X_A}^{-}(\omega)} \right]_{+},
\end{align}
where $[Z(\omega)]_{+}$ is the causal component of $Z(\omega)$, and $S^\pm_{X_AX_A}(\omega)$ is explicitly defined by Eq.~(\ref{SSS}) in Appendix A.
Here, "causal" means considering positive dissipation to guarantee the physically correct behavior of damped oscillation.
Conversely, "non-causal" means negative dissipation which implies physically incorrect behavior of diverging amplitude.

We can determine these Wiener filters as described in Appendix \ref{appFilter}, 
and we find that the Wiener filters are written as
\begin{align}
H_{\delta \x}(\omega)
&=
\frac{1}{S_{X_A X_A}^{+}(\omega)} \biggl[\frac{S_{X_A \delta \x}(\omega)}{S_{X_A X_A}^{-}(\omega)}\biggr]_+=\frac{G_{\delta \x}(\omega)}{S_{X_A X_A}^{+}(\omega)}, 
\quad
G_{\delta \x}(\omega)
=
\frac{C_{1 X_A \delta \x}}{\sqrt{C_{1X_A X_A}}} \frac{\tilde{E} \omega^3+\tilde{F} \omega^2+\tilde{G} \omega+\tilde{H}}{F(\omega)}, 
\label{HGdq}
\\
H_{\delta p}(\omega)
&=
\frac{1}{S_{X_A X_A}^{+}(\omega)} \biggl[\frac{S_{X_A \delta p}(\omega)}{S_{X_A X_A}^{-}(\omega)}\biggr]_+=\frac{G_{\delta p}(\omega)}{S_{X_A X_A}^{+}(\omega)}, 
\quad
G_{\delta p}(\omega)
=
\frac{C_{1 X_A \delta \x}}{\sqrt{C_{1X_A X_A}}} \frac{\tilde{I} \omega^3+\tilde{J} \omega^2+\tilde{K} \omega+\tilde{L}}{F(\omega)}, 
\label{HGdp}
 \\
H_{\delta \Phi}(\omega)
&=
\frac{1}{S_{X_A X_A}^{+}(\omega)} \biggl[\frac{S_{X_A \delta \Phi}(\omega)}{S_{X_A X_A}^{-}(\omega)}\biggr]_+=\frac{G_{\delta \Phi}(\omega)}{S_{X_A X_A}^{+}(\omega)}, 
\quad
G_{\delta \Phi}(\omega)
=
\frac{C_{1 X_A \delta \x}}{\sqrt{C_{1X_A X_A}}} \frac{\tilde{M} \omega^3+\tilde{N} \omega^2+\tilde{O} \omega+\tilde{P}}{F(\omega)}, 
\label{HGdP}
\\
H_{\delta \Pi_{\Phi}}(\omega)
&=
\frac{1}{S_{X_A X_A}^{+}(\omega)} \biggl[\frac{S_{X_A \delta \Pi_{\Phi}}(\omega)}{S_{X_A X_A}^{-}(\omega)}\biggr]_+=\frac{G_{\delta \Pi_{\Phi}}(\omega)}{S_{X_A X_A}^{+}(\omega)}, 
\quad
G_{\delta \Pi_{\Phi}}(\omega)
=
\frac{C_{1 X_A \delta \x}}{\sqrt{C_{1X_A X_A}}} \frac{\tilde{Q} \omega^3+\tilde{R} \omega^2+\tilde{S} \omega+\tilde{T}}{F(\omega)},
\label{HGdPP}
\end{align}
where the explicit formulas necessary for the above 
quantities are defined in Appendix \ref{appFilter}. 
 
\subsection{Conditional covariance}
\label{conditionalc}

Conditional covariance is the squared expected value of the difference between the true value and the physical quantity estimated from the measured output light $X_A$ when the output light $X_A$ is measured, which we write as
\begin{eqnarray}
V_{r_1r_2}&=&\langle (r_1(\omega)-r_1{}_{\text{e}}(\omega))^{\dag} (r_2(\omega)-r_2{}_{\text{e}}(\omega))+(r_{2}(\omega)-r_2{}_{\rm e}(\omega))(r_1(\omega)-r_1{}_{\rm e}(\omega))^{\dag} \rangle.
\end{eqnarray}
Here $r_1,~r_2$ denote the variables $\delta q$, $\delta p$, $\delta\Phi$, and $\delta\Pi_\Phi$, while $r_1{}_{\rm e}, ~r_2{}_{\rm e}$ correspond to the corresponding estimated values, 
which are given with the use of the Wiener filter for each variable, as defined by Eq.~(\ref{dqe}) for $\delta q$ for example, as
\begin{eqnarray}
r_j{}_{\rm e}(\omega)=H_{r_j}(\omega)X_A(\omega)
\end{eqnarray}
with $j=1,~2$. Here $H_{r_j}(\omega)$ denotes $H_{\delta q}(\omega)$, $H_{\delta p}(\omega)$, $H_{\delta \Phi}(\omega)$,  $H_{\delta \Pi_\Phi}(\omega)$, which are given by Eqs. (\ref{HGdq}), (\ref{HGdp}), (\ref{HGdP}), (\ref{HGdPP}),  respectively.

Conditional variance is a quantitative measure of the uncertainty of variables. We introduce the matrices $\bm{V}_c,\bm{V}_r$, which represent the conditional covariance of the pendulum and rotational modes: 
\begin{eqnarray}
\bm{V}_\text{c}=
\begin{pmatrix}
V_{\delta \x \delta \x} & V_{\delta \x \delta p} \\
	V_{\delta \x \delta p} & V_{\delta p \delta p}
\end{pmatrix},
\quad
		\bm{V}_\text{r}=
		\begin{pmatrix}
			V_{\delta \Phi \delta \Phi} & V_{\delta \Phi \delta \Pi_{\Phi}} \\
			V_{\delta \Phi \delta \Pi_{\Phi}} & V_{\delta \Pi_{\Phi} \delta \Pi_{\Phi}}
		\end{pmatrix}.
		\label{VcVr}
\end{eqnarray}
Following the above definition, the components of the conditional covariance are expressed as follows: 
\begin{align}
		V_{\delta \x \delta \x}
		&=
		\frac{1}{2\pi} \int_{-\infty}^{\infty} \mathrm{Re}[S_{\delta \x \delta \x}(\omega) -|G_{\delta \x}(\omega)|^2] d \omega, 
		\label{Vqq}
		\\
		V_{\delta p \delta p}
		&=
		\frac{1}{2\pi} \int_{-\infty}^{\infty} \mathrm{Re}[S_{\delta p \delta p}(\omega) -|G_{\delta p}(\omega)|^2] d \omega, 
		\label{Vpp}
		\\ 
		V_{\delta \x \delta p}
		&=
		\frac{1}{2\pi} \int_{-\infty}^{\infty} \mathrm{Re} [S_{\delta \x \delta p}(\omega) -G_{\delta \x}(\omega)^* G_{\delta p}(\omega)] d \omega, 
		\label{Vqp}
		\\ 
		V_{\delta \Phi \delta \Phi}
		&=
		\frac{1}{2\pi} \int_{-\infty}^{\infty} \mathrm{Re}[S_{\delta \Phi \delta \Phi}(\omega) -|G_{\delta \Phi}(\omega)|^2] d \omega, 
		\label{VPhiPhi}
		\\ 
		V_{\delta \Pi_{\Phi} \delta \Pi_{\Phi} }
		&=
		\frac{1}{2\pi} \int_{-\infty}^{\infty} \mathrm{Re}[S_{\delta \Pi_{\Phi}  \delta \Pi_{\Phi} }(\omega) -|G_{\delta \Pi_{\Phi} }(\omega)|^2] d \omega, 
		\label{VPiPi}
		\\ 
		V_{\delta \Phi \delta \Pi_{\Phi} }
		&=
		\frac{1}{2\pi} \int_{-\infty}^{\infty} \mathrm{Re}[S_{\delta \Phi \delta \Pi_{\Phi} }(\omega) -G_{\delta \Phi}(\omega)^* G_{\delta \Pi_{\Phi} }(\omega)] d \omega.
		\label{VPhiPi}
	\end{align}
From Eqs.~(\ref{sol_dx_o}) and (\ref{sol_dPhi_o}), we have the spectral densities, 
    \begin{align}
    		&S_{\delta \x \delta \x}(\omega)
		=
		\frac{1}{|F(\omega)|^2} 
		(\omega^2-\Delta_{BR}^2)^2\left\{ \frac{2 \kappa \hbar^2 G_0^2 n_c}{M^2 (\kappa^2+\Delta^2)}\left(\kappa^2 \langle \A_{\mathrm{in}}^2(\omega) \rangle+\Delta^2 \langle \B_{\mathrm{in}}^2(\omega) \rangle \right) + \frac{\langle \xi_m^2(\omega) \rangle}{M^2} \right\},  \\
		&S_{\delta \Phi \delta \Phi}(\omega)
		=
		\frac{1}{|F(\omega)|^2} 
		(\omega_{BR}^2)^2 \left\{ \frac{2 \kappa \hbar^2 G_0^2 n_c}{M^2 (\kappa^2+\Delta^2)^2}\left(\kappa^2 \langle \A_{\mathrm{in}}^2(\omega) \rangle+\Delta^2 \langle \B_{\mathrm{in}}^2(\omega) \rangle \right) + \frac{\langle \xi_m^2(\omega) \rangle}{M^2} \right\},
	\end{align}
    and the other components
	\begin{align}
		&S_{\delta p \delta p}(\omega)
		=
		M^2 \omega^2 S_{\delta \x \delta \x}(\omega), 
		\quad
		S_{\delta \x \delta p}(\omega)
		=
		i M \omega S_{\delta \x \delta \x} (\omega), 
		\quad
		\\
		&S_{\delta \Pi_{\Phi} \delta \Pi_{\Phi}}(\omega)
		=
		J^2 \omega^2 S_{\delta \Phi \delta \Phi}(\omega), 
		\quad
		S_{\delta \Phi \delta \Pi_{\Phi}}(\omega)
		=
		i J \omega S_{\delta \Phi \delta \Phi} (\omega).
\end{align}

The integrals for the conditional variances are calculated using the residue theorem.
The residue theorem uses the value of a pole of the integrand, which is obtained as $|F(\omega)|^2=0$.
Two types of poles appear from $|F(\omega)|^2=0$, which corresponds to 
pendulum, and rotational modes.
The perturbation $\delta q$ and its conjugate momentum $\delta p$ resonate at the frequency of the pendulum mode and the effect of the rotational mode is very small.
Therefore, we can treat the dissipation $\Gamma_{\text{r}}(\omega)$ that is proportional to the inverse of the frequency as $\Gamma_{\text{r}}(\omega)\simeq \Gamma_{\text{r}}(\Omega) \Omega/\omega$.
For the fluctuation $\xi_m$ expressed by Eq.~(\ref{xi_m^2_avg}) or Eq.~(\ref{xi_m^2_avgb}), we have $k_B T_0'/\hbar \simeq 10^{10} [1/\mathrm{s}]$ even for very low temperatures $T_0'\sim 1 \mathrm{K}$. Then, the following approximation holds: 
	\begin{align}
		M \hbar \gamma_m \omega \coth \left(\frac{\hbar \omega \gamma_m}{2 k_B T_0 \Gamma_{\text{r}}(\omega)} \right) 
		\simeq M \hbar \omega \gamma_m  \frac{2 k_B T_0{\Gamma_{\text{r}}(\omega)}}{\hbar \omega {\gamma_m}} 
		\simeq  M \hbar \Omega \gamma_m (2 \bar{n}_p+1),
	\end{align}
where we define 
	\begin{align}
		\bar{n}_\text{p}
		=
		\left(\exp \left\{\frac{\hbar \Omega{\gamma_\text{m}}}{k_\text{B} T_0{\Gamma_{\text{r}}(\omega)}} \right\}-1\right)^{-1} \simeq \frac{k_\text{B} T_0{\Gamma_{\text{r}}(\omega)}}{\hbar \Omega {\gamma_\text{m}}}-\frac{1}{2}.
	\end{align}
	Here, $\bar{n}_\text{p}$ denotes the average {phonon} number of thermally excited states.

\subsection{Comparison with the point-particle mirror model}
In the previous analysis \cite{MY}, the mirror was modeled as a point particle, where the rotational mode is neglected and only the pendulum mode is relevant. In Appendix \ref{pointone}, we presented a review of the Wiener filter and the conditional variance for a point-particle mirror model, 
Eq.~(\ref{Vcm}) with Eqs.~(\ref{Vcmqq})~$\sim$~(\ref{Vcmqp}), 
which was developed in Ref.~\cite{MY}.
The point particle model is quite simple compared with the two-mode theory developed in the present paper, and it is useful to investigate whether the prediction of the two-mode theory is reproduced with the simple model or not. 
In order to investigate this point, we consider the following two models on the basis of the Wiener filter of the one-mode theory. 

Modified model I: In Appendix~\ref{Appenfour},
we have found the conditional variance  ${\bm V_c'}$ when the Wiener filter on the basis of the one-mode theory is adopted for the output light $X_A(\omega)$ obtained in the beam model Eqs.~(\ref{V'_c}), 
whose components are given by 
Eqs.~(\ref{V'_dqdq})~$\sim$~(\ref{V'_dqdp}),
where $H_{\delta q_m}(\omega)$ and $H_{\delta p_m}(\omega)$ are the Wiener filters of the one-mode theory developed in Appendix \ref{onemodeWF}.
With the use of this conditional covariance, we can evaluate the influence of the rotational mode on the conditional variance developed in the previous section. 

Modified model II: In the previous paper \cite{Sugiyama}, we showed that
the quality factor of the pendulum mode was reduced by a factor of four when the pendulum and rotational modes resonate with each other based on the analogy of a coupled harmonic oscillator. This fact inspires the model of the one-mode Wiener filter replacing the dissipation $\Gamma$ with $N\Gamma$, where adopt $N=4.18$. 
Explicitly, the modified model II is the same as the modified model I but with the dissipation 
replaced with $N\Gamma$.

\subsection{Results}
 In our analysis, the theoretical prediction
 is based on the model parameters listed in Table I, which were referred to in the previous works 
 \cite{MY,Lopez,Matsumoto,Sugiyama}.
 Especially, we assume the temperature $T_0=300$K, then
 the effective temperature of the mirror is reduced to $T_{\rm eff}=T_0\Gamma_r/\gamma_m\simeq 7.5\times 10^{-2}$~K due to the feedback control in our model. 
 Furthermore, due to the Wiener filtering process, the noises are reduced at the level that the high purity of $0.5$ will be achieved by using the two-mode Wiener filter consistently for the two-mode model, while it could be worse when we use one-mode Wiener filter for the two-mode model.
		To demonstrate the results, we introduce normalized conditional variances by multiplying the following dimensional quantities, which are dimensionless quantities normalized to unity for the vacuum ground state: 
	\begin{align}
		&V_{\delta \x \delta \x} \times \frac{2 M \omega_{AR}}{\hbar}, 
		\quad
		V_{\delta p \delta p} \times \frac{2}{\hbar  M \omega_{AR}}, 
		\quad
		V_{\delta \x \delta p} \times \frac{2}{\hbar}, 
		\quad
		\\
		&V_{\delta \Phi \delta \Phi} \times \frac{2J \Delta_{BR}}{\hbar},
		\quad
		V_{\delta \Pi_{\phi} \delta \Pi_{\phi}} \times \frac{2}{\hbar J \Delta_{BR}},
		\quad
		V_{\delta \Phi \delta \Pi_{\Phi}} \times \frac{2}{\hbar}.
	\end{align}
	\begin{figure}[H]
		\centering
		\includegraphics[width=105mm]{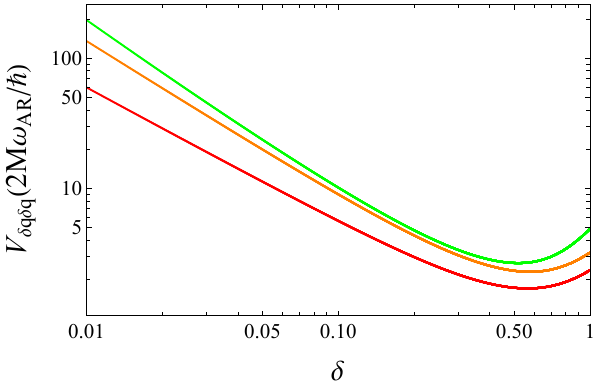}
		\caption{
			Dimensionless variance of $\delta\x$ as a function of the normalized detuning $\delta$, 
which is normalized to unity for the vacuum state by the factor $2M\omega_{\rm AR}/\hbar$. 
The red curve is the result of the Wiener filter of the two-mode theory, $V_{\delta \x\delta\x}(2M\omega_{AR}/\hbar)$ with Eq.~(\ref{Vqq}), while  the green curve is the result of the one-mode Wiener filter of the pendulum mode alone in the limit of the point mass mirror (Modified model II), $V'_{\delta q\delta q}(2M\omega_{AR}/\hbar)$ with Eq.~(\ref{V'_dqdq}). The orange curve shows the result with the one-mode Wiener filter with the dissipation $4.18\Gamma$ instead of $\Gamma$ (Modified model II). See Appendix \ref{point} for further details of the one-mode Wiener filter.
When the detuning is small, the information of the position $\delta q$ cannot be sufficiently obtained by the measurement of $X_A$ as the detuning determines the efficiency of the measurement by Eq.(\ref{dxdqnoise}). Then, the variance of $\delta q$ is large when the detuning is small, and the variance reduces for higher detuning, but further higher detuning prevents reducing the noise because it
hinders efficient measurement of the oscillator's position
for $\delta \simgt 0.5$. Therefore there is a suitable value of the detuning of the range $0.2\simlt\delta\simlt 0.5$. 
This also explains the behaviors of Figs.~\ref{fig:vpp_G}$~\sim$ \ref{fig:pure_G} because $\delta q$ and $\delta p$ are related to each other by Eq.~(\ref{sol_dp_o}).}
\label{fig:vxx_G}
\end{figure}
\begin{figure}[H]
\centering
\includegraphics[width=105mm]{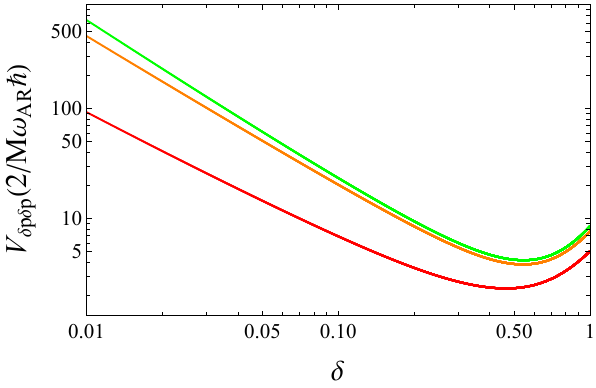}
\caption{
Same as Fig.~\ref{fig:vxx_G} but for the variance of the momentum $\delta p$, which is normalized to unity for the vacuum state by divided by the factor $2/M\omega_{\rm AR}\hbar$. 
The red curve is the result with the Wiener filter of the two-mode theory $V_{\delta p\delta p}(2/M\omega_{AR}\hbar)$ with Eq.~(\ref{Vpp}), while the green curve is the result with the Wiener filter of the one-mode theory $V'_{\delta p\delta p}(2/M\omega_{AR}\hbar)$ with Eq.~(\ref{V'_dpdp}) defined as (Modified model I). The orange curve is the result of the one-mode Wiener filter with the dissipation $4.18\Gamma$ instead of $\Gamma$ (Modified model II).
\label{fig:vpp_G}}
\end{figure}
\begin{figure}[H]
\centering
\includegraphics[width=105mm]{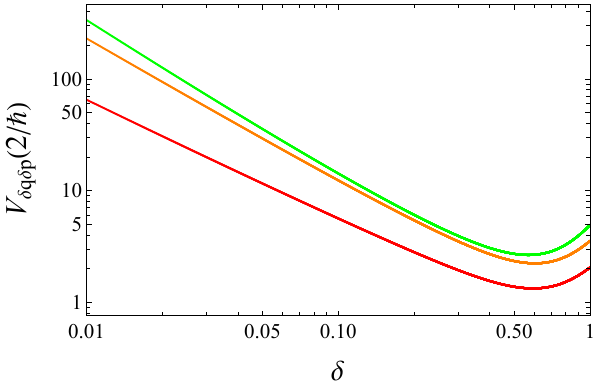}
\caption{
Same as Fig.~\ref{fig:vxx_G} but for the covariance of $\delta q$ and $\delta p$. 
The red curve is  $V_{\delta q\delta p}(2/\hbar)$ with Eq.~(\ref{Vqp}),   while the green curve is $V'_{\delta q\delta p}(2/\hbar)$ with Eq.~(\ref{V'_dqdp}) is the one-mode Wiener filter (Modified model I). The orange curve is the result of the one-mode Wiener filter with the dissipation $4.18\Gamma$ instead of $\Gamma$ (Modified model II). As is explained in the caption of Fig.~\ref{fig:vxx_G}, the efficiency of measurement of $\delta q$ through $X_A$ determines the behaviors $\delta \simlt 0.5$. The large covariance between $\delta q$ and $\delta p$ in the regime of small detuning does not mean a large correlation between them as each variance is large at the same time.
\label{fig:vxp_G}}
\end{figure}
\begin{figure}[H]
\centering
\includegraphics[width=105mm]{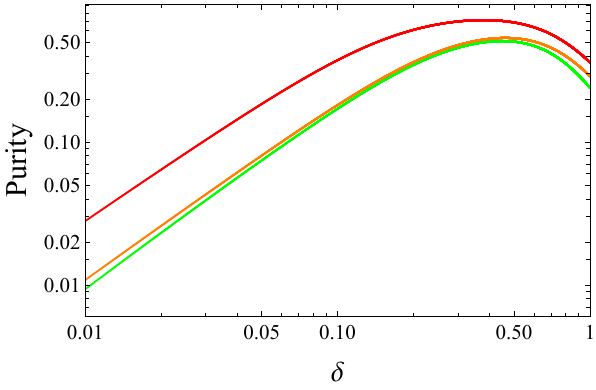}
\caption{
Purity as a function of the detuning $\delta$.
The red  curve is the result of the Wiener filter in the two-mode theory, ($1/\sqrt{\det \bm{V}_c}$), while the green curve is the result of the one-mode Wiener filter of the pendulum mode alone in the limit of the point mass mirror, ($1/\sqrt{\det \bm{V}_c'}$), (Modified model I). The orange curve is the result of the one-mode Wiener filter with the dissipation $4.18\Gamma$ instead of $\Gamma$ (Modified model II).
			\label{fig:pure_G}}
	\end{figure}
\begin{figure}[H]
\centering
\includegraphics[width=75mm]{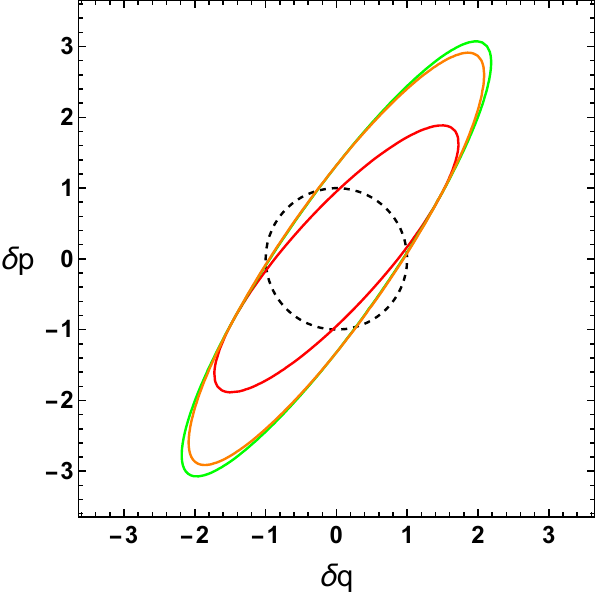}
\caption{
Contour for the Wigner function satisfying $W(\delta q, \delta p) = e^{-1}W_{\rm max}$ (Wigner ellipse) on the plane of $\delta q$ and $\delta p$ with the covariance with the Wiener filter of the two-mode theory $\bm V_c$ (red ellipse).  
The green ellipse shows the same but 
with that of the one-mode modeling $\bm V_c'$ (Modified model I).
The orange ellipse is the result of the one-mode Wiener filter with the dissipation of $4.18\Gamma$ instead of $\Gamma$ (Modified model II).
The green and red ellipses are the results of the modified model I and II, respectively.
Here the Wigner function is defined as $W(\delta q,\delta p)=~ 2/(\pi\sqrt{{\rm det}V})\exp[-{\bm r}^{\rm T}\bm V^{-1}\bm r]$, where ${\bm r}=(\delta q,\delta p)$, and $\bm V$ denotes $\bm V_c$ or $\bm V_c'$.
The detuning is fixed as $\delta=0.2$. The black dashed circle represents the vacuum state.
\label{fig:vxp_phase_G}}
\end{figure}

Fig.\ref{fig:vxx_G} plots the dimensionless variance of $\delta \x$ as a function of the normalized detuning $\delta$, where the red curve represents the result with the Wiener filter of the two-mode theory developed in the previous subsection, while the green curve is the result with the Wiener filter of the one-mode model of the pendulum mode alone in the limit of the point mass mirror (Modified model I). In Appendix \ref{point} we described the details of the one-mode model.
The orange curve is the result of applying the Wiener filter to one-mode theory when the dissipative decay rate $\Gamma$ is replaced by $4.18\Gamma$ (Modified model II), which is motivated by our finding that the quality factor of the pendulum mode is reduced by the factor four due to the presence of the rotational model~ \cite{Sugiyama}.
Similarly, Fig.~\ref{fig:vpp_G} plots the dimensionless variance of the momentum $\delta p$, while Fig.~\ref{fig:vxp_G} shows the dimensionless covariance of $\delta \x$ and momentum $\delta p$.
 The behaviors of these curves as a function of 
 detuning $\delta$ is understood as the efficiency to determine $\delta q$ through a measurement of $X_A$, which is specified by Eq.~(\ref{dxdqnoise}).

As is described in the Appendix \ref{point}, the one-mode model describes the mirror as a mass point, which is adopted in the previous study in Ref.~\cite{Matsumoto}. Therefore, the difference between the red and green curves can 
be understood as a finite-size effect of the mirror in the present model. The results demonstrate that proper treatment of the rotational model is important for achieving a quantum state close to the vacuum state. This is illustrated in Fig.~\ref{fig:pure_G}, which plots
the purity ($=1/\sqrt{\det \bm{V}_c}$). Purity is $1$ for the pure state. 
The same color for each curve in Fig.~\ref{fig:pure_G} and Fig.~\ref{fig:vxx_G} adopts the same model.
The red curve is always larger than the green curve for $\delta< 1$.
Therefore, Fig.~\ref{fig:pure_G} shows that neglecting the rotational mode prevents the
state of the mirror from generating a quantum-squeezed state. 
This is illustrated in Figs.~\ref{fig:vxp_phase_G}, which plots the 
Wigner ellipse, that is, the contour of the Wigner distribution function in phase space
for the states whose covariance are: ${\bm V}_c$ as defined by Eq.~(\ref{VcVr}) (red curve), 
${\bm V}_c'$ as defined  Eq.~(\ref{V'_c}) (green curve, Modified model I), and ${\bm V}_c'$ as defined by Eq.~(\ref{V'_c}) with 
$4.18\Gamma$ instead of $\Gamma$ (orange curve, Modified model II). The normalized detuning is fixed at $\delta=0.2$. 
The same color for each curve in Fig.~\ref{fig:vxp_phase_G} and Fig.~\ref{fig:vxx_G} corresponds to the same model.
The black dashed circle represents a Wigner ellipse for the vacuum state. 
Fig.~\ref{fig:vxp_phase_G} demonstrates that two-mode modeling plays an important role in generating a quantum squeezed state.

\begin{figure}[t]
\centering
\includegraphics[width=8.5cm]{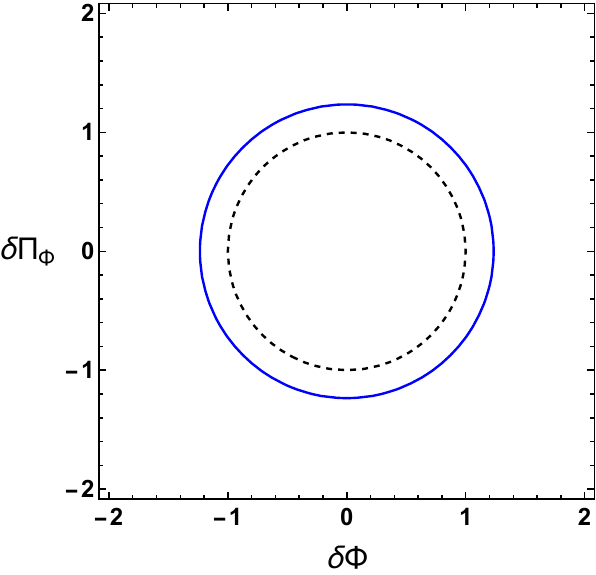}
\caption{
Wigner ellipse in the phase space of $\delta \Phi$ and $\delta \Pi_\Phi$ with the covariance ${\bm V_r}$ (blue curve). The parameters in this case are the same as those in Fig.~\ref{fig:vxp_phase_G}. The black dashed circle represents the vacuum state.
\label{fig:vPPi_phase_G}}
\end{figure}

Figures {\ref{fig:vxx_G}}$\sim${\ref{fig:vxp_G}} show that the orange curve has smaller values than the green curve. 
This indicates that $\Gamma$ can be used as a parameter to improve the Wiener filter process within the one-mode theory. 
This is demonstrated in Figures {\ref{fig:pure_G}} and {\ref{fig:vxp_phase_G}} as
the purity and Wigner ellipse in orange are closer to the results of the two-mode theory shown in red. 
However, the improvement is limited, and simply changing the dissipative $\Gamma$ of the one-mode Wiener filter does not completely account for the effects of the rotation mode of the two-mode Wiener filter.

Finally in this section, we mention the state of the rotational mode. Fig.~\ref{fig:vPPi_phase_G} shows the Wigner ellipse in the phase space of the rotational modes $\delta \Phi$ and $\delta \Pi_\Phi$. 
It is obtained using the covariance ${\bm V}_r$ defined by Eq.~(\ref{VcVr}), where the parameters are the same as those in Fig.~\ref{fig:vxp_phase_G}.
The state of the rotational mode shown in Fig.~\ref{fig:vPPi_phase_G} is close to that
in the vacuum state, and the Wigner ellipse is approximately circular. 
This implies that the rotational mode is also cooled similarly to the pendulum mode, but this state is not squeezed. This result might be regarded as the equipartition law of energy for the mode coupled to the degrees of freedom of measurement cooling.
However, these results for the rotational mode should be carefully considered because we only consider noise associated with the internal friction of the beam, and neglected the other noise. 

\section{Discussions}
\label{secdisc}
\subsection{One mode filter with the constrained range in Fourier space}

\begin{figure}[t]
\begin{center}\includegraphics[width=110mm]{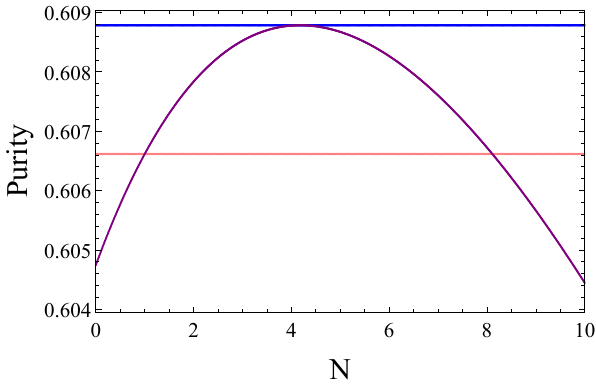}
\caption{
The purple curve shows the purity using the one-mode Wiener filter (modified model III) by replacing the dissipation $\Gamma$ with $N\Gamma$ when
the contribution from the pole of the rotational mode is omitted. Here we fixed at $\delta=\delta_0(=0.2)$.
The blue and pink lines are plotted for comparison.
The blue line is the result using the Wiener filter with the two-mode theory, while
the pink line is the result using the one-mode Wiener filter with the mirror treated as a mass point $N=1$ (modified model III).
The contribution from the pole of the rotational mode is discarded in all the cases.
\label{fig:purity_N11}}
\end{center}
\end{figure}

We have discussed the theoretical aspects, but an analysis considering actual experiments is also necessary. In the analysis of the spectral density with the Wiener filter, the data in the Fourier space is integrated to calculate the conditional variance.
Thus, it is normal to restrict the frequency range of the Fourier space
relevant to the measurement without using the entire frequency space in actual experiments.
Since we focus on the pendulum mode of the mirror, the measurement is considered to be 
limited to the finite region of frequency including the pendulum mode.
For example, our model should include the range of  the frequency of pendulum mode in the range of $100~{\rm Hz} <\omega/2\pi < 700~{\rm Hz}$, while that of the rotational mode is $\omega/2\pi\simeq 27 $~Hz.
In other words, we can exclude the other range of Fourier space when conducting this experimental analysis. In this section, we consider the case excluding the frequency domain of the mirror's rotation mode $\omega/2\pi \approx 27 \text{Hz}$.


In order to take this effect into account, we restrict the range of integration concerning $\omega$ in estimating the conditional variance Eqs.~(\ref{Vqq})$\sim$(\ref{VPhiPhi}). 
As discussed in Sec.~\ref{conditionalc}, the integrals for the conditional variances are calculated using the residue theorem.
The residue theorem uses the values of the poles of the integrand, obtained from $|F(\omega)|^2=0$. Two types of poles appear from $|F(\omega)|^2=0$, which correspond to pendulum, and rotational modes, where the real part of the pole of the rotational modes is much smaller than that of the pendulum mode. 
Here we should consider the case that the integration concerning $\omega$ is performed only around the pendulum mode.
We assume that this case is described by evaluating the residue of the pole corresponding to the pendulum mode of the integrand, and by discarding the residue of the pole corresponding to the rotation mode. 
Namely, we evaluate the conditional covariance $\bm V_c'$ in Eqs.~(\ref{V'_dqdq}), 
$\sim$(\ref{V'_dpdp}) but with neglecting the contribution of the poles of the rotational mode.
We demonstrate that this provides us with a simple and useful method with the one-mode Wiener filter. 
We call this model the modified model III.  


Fig.~{\ref{fig:purity_N11}} compares the purity with the 
Wiener filter of the two-mode theory (blue line), 
Modified model III (pink line), and Modified model III with the dissipation $N\Gamma$ as a function $N$ (purple curve). 
In all these models, the contribution of the pole from the rotational mode is discarded.
We note that the blue line and the pink line, which are defined only at $N=1$, are plotted for a comparison with the purple curve, which is defined as a function of $N$.
The result of the modified model III well reproduces the covariance matrix with the exact two-mode Wiener filter $\bm V_c$ within $10^{-6}~\%$ for $\delta \leq 1$.
This difference can be smaller by changing the dissipation rate $\Gamma$ as shown in Fig.~{\ref{fig:purity_N11}}. 
The purity of the modified model III with the dissipation $N\Gamma$ takes the maximum value 
and coincides with the two-mode theory for $N\simeq 4$, which is explained by the fact that the quality factor of the pendulum mode is reduced by factor four due to the presence of the rotational mode \cite{Sugiyama}.
This suggests that the one-mode Wiener filter analysis can provide a good approximation by removing the low-frequency Fourier modes of the Wiener filter. Namely, one can use the one-mode Wiener filter by properly choosing the range of the modes in the Fourier space discarding the low frequency Fourier modes $\omega/2\pi\simlt 30 \sim 100 $~Hz. 
This is guaranteed by the fact that the frequency of the pendulum and
rotational modes are quite different from each other in our model's parameters.

\begin{figure}[b]
  \begin{center}
  \includegraphics[width=150mm]{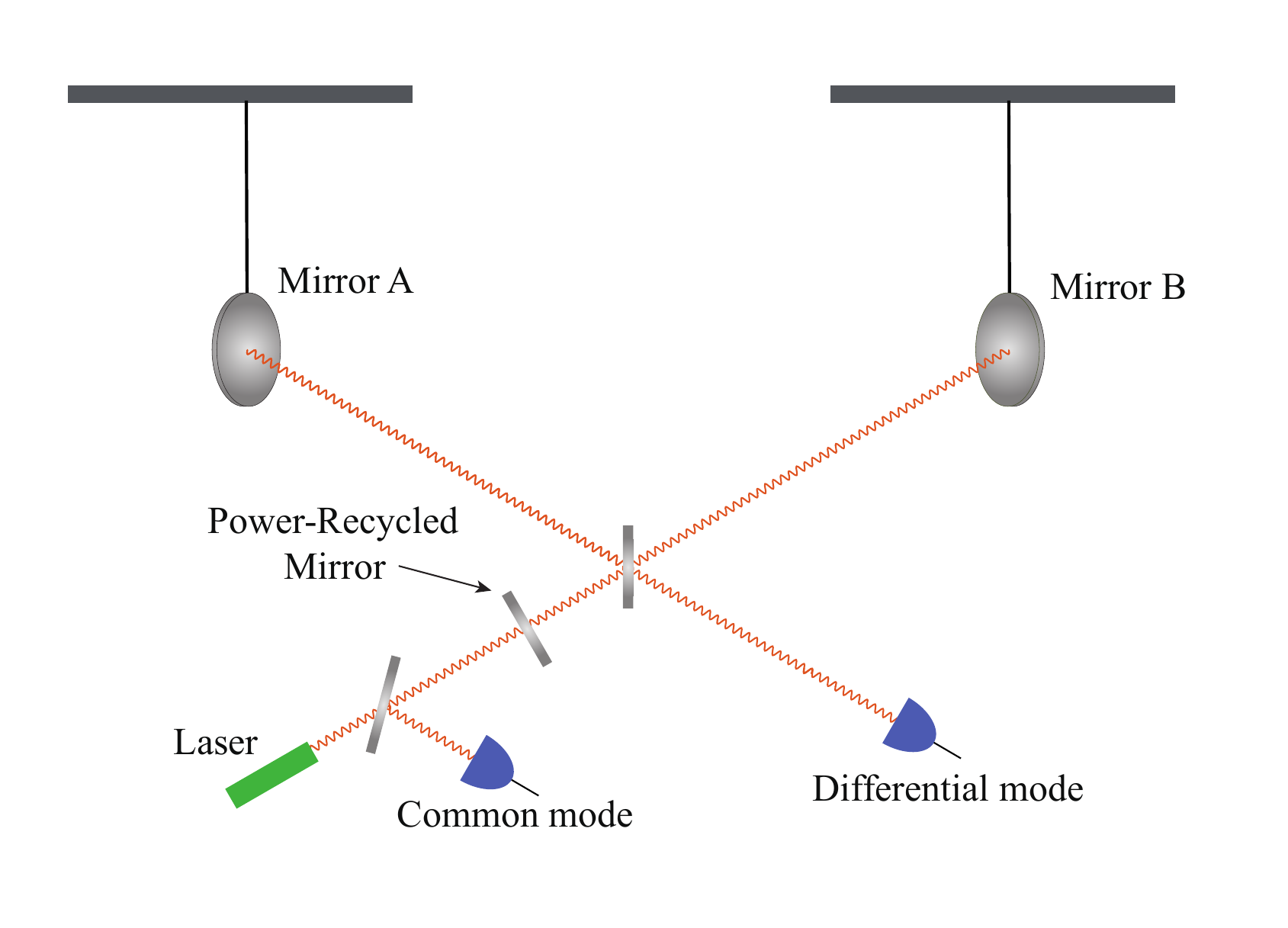}\caption{
Schematic plot of the Power-Recycled–Fabry-Perot-Michelson interferometer, which is assumed to compute the entanglement between two mirrors.
\label{fig:setup}}\end{center}
\end{figure}

\subsection{Implication for generating entanglement}
Finally, we consider the implication of the above results for generating the quantum entanglement through a half mirror between two $\text{mg}$ scale mirrors that make up a power-recycled-Fabry-Perot Michelson interferometer, demonstrated in Ref.~\cite{Miki3}. 
See figure \ref{fig:setup} for a schematic plot of our setup to be considered in the present paper. 
The generation of entanglement between macroscopic objects is a frontier of modern physics to explore the boundary of the quantum world and classical world \cite{Whittle}. 
The authors of Ref.~\cite{Miki3} investigated the conditions of generating entanglement between the
$7~\text{mg}$ mirrors consisting of a Fabry-Perot-Michelson interferometer, where each mirror is modeled within the one-mode theory.
The impact of the two-mode theory on the feasibility of generating entanglement is of interest. 
The authors of Ref.~\cite{Miki3} showed that both the purity and squeezing of the mechanical common and differential modes are required for the entanglement generation, which is optimized at the detuning of the differential mode $\delta_{-}=-\Delta/2\kappa_{-}\sim0.2$.
 Then, the purity of the differential mode $P_{-}\simgt0.5$ roughly needs to generate the entanglement for the parameters in Table \ref{tab:parameter}.
 The purity of the differential mode is regarded as the purity in Fig~\ref{fig:pure_G}. We obtain the purity of the common mode by only reducing the optical decay rate by the power-recycling mirror.
  In Fig.~\ref{fig:pure_G}, the purity with the one-mode filter (green curve) is smaller than $0.5$, but the purity with the two-mode filter (red curve) is larger than the required value.
  
\begin{figure}[tbp]
\begin{center}\includegraphics[width=110mm]{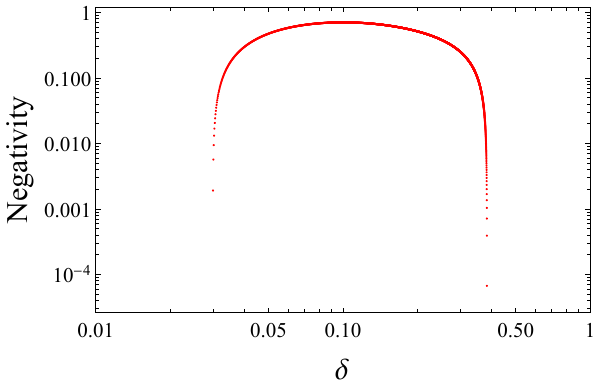}
\caption{
Logarithmic negativity generated between the two mirrors as a function of the normalized detuning $\delta$. 
The red marks show the result with the exact two-mode Wiener filter.
We note that the result with the one-mode Wiener filter
but with neglecting the contribution from the pole of the rotational modes 
in evaluating $\bm V_c'$ well overlaps the result of the red marks, which means that the 
one-mode Wiener filter by properly choosing the range of the modes in the Fourier space
reproduces the result of the two-mode Wiener filter. 
\label{fig:NN2}}
\end{center}
\end{figure}

 We focus on the logarithmic negativity between the pendulum mode of the two mirrors, as in Ref.~\cite{Miki3}. 
The logarithmic negativity is an indicator of entanglement between the two mirrors, which is generated due to the asymmetric property
between the common mode and the differential mode of the two mirrors that consist of the Fabry-Perot-Michelson interferometer.  The logarithmic negativity  can be computed with the covariance matrix of the two mirrors. (See Eq.(63) in Ref.\cite{Miki3} for the definition). 
Figure \ref{fig:NN2} plots the logarithmic negativity between the two mirrors
as a function of the normalized detuning, where we assumed the 
parameter of the ratio of the dissipation of the differential mode to
the common mode $\kappa_-/\kappa_+=3$ with $\kappa_-(=\kappa)$ and $\delta_{-}=-\Delta/2\kappa_{-}(=\delta)$.
Fig.~\ref{fig:NN2} demonstrates that the logarithmic  negativity is positive
for $0.03<\delta<0.33$, indicating the entanglement generation between the two mirrors.
This can be achieved when we use the exact two-mode Wiener filter or the Modified model III.

\section{Summary and Conclusion\label{summary}}
We developed a beam model~\cite{Saulson} coupled with a cavity light to explore a precise theoretical model of an optomechanical system that generates a quantum state of macroscopic objects for testing the quantum nature of gravity. 
This is useful for describing the quantum fluctuations in the motion of a suspended mirror under continuous measurement and feedback control as an extended model of the optomechanical oscillator system analyzed in Ref.~\cite{Matsumoto}.
We developed a two-mode theory for the mirror described by the pendulum and rotational modes coupled with each other. Some aspects of the usefulness of the two-mode theory were discussed in Ref.~\cite{Sugiyama} by demonstrating that the model reproduces the quality factor, which is consistent with the experimental results in Ref.~\cite{Matsumoto}. 
In the present study, we explicitly demonstrated that the structural damping dissipation was derived from the equation of motion for the pendulum mode of the mirror due to the internal dissipation of the beam. 
We then introduced thermal noises for the pendulum mode of the mirror, which were consistent with those predicted by the internal friction of the beam, to satisfy the fluctuation-dissipation relation, and clarified the quantum state that can be realized through continuous measurement and feedback control. 
We focused on constructing a Wiener filter for the two-mode theory of the pendulum and rotational modes. We utilized this to investigate the quantum state by evaluating the conditional covariance matrix using Wiener filter analysis.
We found that the quantum-squeezed state can be generated using the Wiener filter for the two-mode theory with the parameters from the state-of-the-art technique, and this is expected in the near future.
In other words, the correctly filtered covariance was squeezed to be smaller than the vacuum fluctuations.

We also demonstrate that the covariance with the different Wiener filters with the one-mode model of the mirror, which corresponds to the limit of the point mass of the mirror, 
becomes larger than that predicted by the two-mode model.
We showed that the two-mode filter achieves the purity required for entanglement generation between two mirrors that make up an interferometer in Ref.~\cite{Miki3}, but a simple analysis with the one-mode filter cannot. We also investigated whether simply changing the dissipation $\Gamma$ of the one-mode theory can improve the covariance with the Wiener filter of the two-mode theory with the rotational mode and found that changing the dissipation $\Gamma$ of the one-mode theory did not explain the mirror rotation mode.
However, we also investigated one-mode Wiener filter analysis while discarding the contribution from the pole of the rotational mode,
which well reproduced the result of the two-mode Wiener filter.  
This suggests that properly choosing the range of the modes in the Fourier space can provide a good approximation to reproduce
the result of the two-mode Wiener filter when the frequency of the pendulum and the rotational modes are well separated.
These results demonstrate that the rotational mode could hinder the generation of the quantum state of the pendulum mode with continuous measurements and feedback control unless additional degrees of freedom are adequately considered in the theoretical predictions.

	\acknowledgements
	We thank Naoki Yamamoto and Yuta Michimura for the discussions and comments on this topic in the present work.
	Y.S. was supported by a Kyushu University Innovator Fellowship in Quantum Science.
	K.Y. was supported by JSPS KAKENHI, Grant No.~JP22H05263 and No.~JP23H01175.
	N.M. was supported by JSPS KAKENHI Grant No.~JP19H00671 and JST FORESTO Grant No.~JPMJFR202X.
	D. M. was supported by JSPS KAKENHI, Grant No.~JP22J21267.
	
	\appendix

 \section{Wiener filter for the two-mode theory}
 \label{appFilter}
	
By calculating the spectral density of $X_A(\omega)$ using Eq.~(\ref{XA_sol_noise}), 
we obtain
	\begin{align}
		S_{X_A X_A}(\omega)
		=
		C_{1X_A X_A} \frac{J(\omega)}{|F(\omega)|^2} \label{S_XAXA},  
	\end{align}
where we defined
	\begin{align}
		J(\omega)
		=
		|F(\omega)|^2 +\frac{C_{2X_AX_A}}{C_{1X_AX_A}} (\omega^2
		-\Delta_{BR}^{2})
		(F(\omega)^*+F(\omega))+\frac{C_{3X_AX_A}}{C_{1X_AX_A}}(\omega^2-\Delta_{BR}^2)^2, 
	\end{align}
and
	\begin{align}
		C_{1X_AX_A} 
		&=
		\eta \left\{\left(1-\frac{2 \kappa^2}{\kappa^2+\Delta^2} \right)^2\langle \A_{\mathrm{in}}^2 (\omega) \rangle +\left(\frac{2 \kappa \Delta}{\kappa^2+\Delta^2} \right)^2 \langle \B_{\mathrm{in}}^2 (\omega) \rangle \right\}+(1-\eta) \langle \A_{\mathrm{in}}'^2 (\omega) \rangle,  \\
		C_{2X_AX_A} 
		&=
		\eta \frac{4 \hbar G_0^2 n_c \Delta \kappa^2}{M (\kappa^2+\Delta^2)^2} \left\{-\left(1-\frac{2 \kappa^2}{\kappa^2+\Delta^2} \right) \langle \A_{\mathrm{in}}^2 (\omega) \rangle+\frac{2 \Delta^2}{\kappa^2+\Delta^2} \langle \B_{\mathrm{in}}^2 (\omega) \rangle  \right\}, 
  \\
		C_{3X_AX_A} 
		&=
		\eta \left[ \left\{\frac{4 \hbar G_0^2 n_c \Delta \kappa}{M (\kappa^2+\Delta^2)^2} \right\}^2 ( \kappa^2 \langle \A_{\mathrm{in}}^2 (\omega) \rangle+\Delta^2 \langle \B_{\mathrm{in}}^2 (\omega) \rangle)+\left(\frac{2 G_0 \sqrt{n_c} \Delta \sqrt{2\kappa}}{\kappa^2+\Delta^2} \right)^2  \frac{\langle \xi_m^2 (\omega) \rangle}{M^2} \right]. \nonumber 
	\end{align}
	The spectral density $S_{X_A X_A}(\omega)$ consists of the causal $S_{X_A X_A}^+(\omega)$ and non-causal $S_{X_A X_A}^-(\omega)$ parts, which are expressed as 
	\begin{align}
		S_{X_A X_A}(\omega) 
		=
		S_{X_A X_A}^+(\omega) S_{X_A X_A}^-(\omega), 
		\quad
		S_{X_A X_A}^+(\omega)
		=
		\sqrt{C_{1 X_A X_A}} \frac{F'(\omega)}{F(\omega)}, 
		\quad
		S_{X_A X_A}^-(\omega)
		=
		\sqrt{C_{1 X_A X_A}} \frac{F'(\omega)^*}{F(\omega)^*}. 
  \label{SSS}
	\end{align}
Here $F'(\omega)$ is a causal function expressed as
	\begin{eqnarray}
		F'(\omega)=- \omega^4+\tilde{A} \omega^3 +\tilde{B} \omega^2+ \tilde{C} \omega +\tilde{D}
	\end{eqnarray}
with the coefficients $\tilde{A},\tilde{B},\tilde{C},\tilde{D}$, which are 
determined using the process described below.
In particular, to determine the coefficients $\tilde{A} \sim \tilde{D}$,  we use 
the fact that $F( \omega)'$ is related to its complex conjugate by
	\begin{eqnarray}
		F'(\omega) F'(\omega)^* =J(\omega) \label{J_omega}.
	\end{eqnarray}
We find the coefficients $\tilde{A},\tilde{B},\tilde{C},\tilde{D}$ by comparing the coefficients in Eq.~(\ref{J_omega}) and solving the following equation. 
	\begin{eqnarray}
		\begin{cases}
			{-(\tilde{A}+\tilde{A}^*)=0 ,} \\
			{\tilde{A}\tilde{A}^*-\tilde{B}-\tilde{B}^*= -2\frac{C_{2 X_A X_A}}{C_{1 X_A X_A}}+\gamma_m^2-2 \Delta_{BR}^2-2 \omega_{AR}^2 ,}\\
			{\tilde{A}\tilde{B}^*+\tilde{A}^* \tilde{B}-\tilde{C}-\tilde{C}^*=0 ,}\\
			{\tilde{B} \tilde{B}^*+\tilde{A}\tilde{C}^*+\tilde{A}^* \tilde{C}-\tilde{D}-\tilde{D}^*=\frac{C_{3 X_A X_A}}{C_{1 X_A X_A}}-2 \gamma_m^2 \Delta_{BR}^2+2 \Delta_{BR}^2 \omega_{AR}^2+(\Delta_{BR}^2+\omega_{AR}^2)^2}\\
			{\hspace{145pt}+\frac{C_{2 X_A X_A}}{C_{1 X_A X_A}}(4 \Delta_{BR}^2+2\omega_{AR}^2)-2 \Delta_{AR}^2 \omega_{BR}^2 , }\\
			{\tilde{B} \tilde{C}^* +\tilde{B}^* \tilde{C}+\tilde{A} \tilde{D}^* +\tilde{A}^*\tilde{D}=0 ,}\\
			{\tilde{C} \tilde{C}^*+\tilde{B} \tilde{D}^*+\tilde{B}^* \tilde{D}=-2\frac{C_{3 X_A X_A}}{C_{1 X_A X_A}}\Delta_{BR}^2+\gamma_m^2 \Delta_{BR}^4-2\frac{C_{2 X_A X_A}}{C_{1 X_A X_A}}(\Delta_{BR}^4+2 \Delta_{BR}^2 \omega_{AR}^2 -\Delta_{AR}^2 \omega_{BR}^2)}\\
			{\hspace{100pt}+2(\Delta_{BR}^2+\omega_{AR}^2)(-\Delta_{BR}^2 \omega_{AR}^2+\Delta_{AR}^2 \omega_{BR}^2),}\\
			{\tilde{C} \tilde{D}^*+\tilde{C}^*\tilde{D}=0 ,}\\
			{\tilde{D}\tilde{D}^*=\frac{C_{3 X_A X_A}}{C_{1 X_A X_A}} \Delta_{BR}^4+2 \frac{C_{2 X_A X_A}}{C_{1 X_A X_A}} \Delta_{BR}^2(\Delta_{BR}^2 \omega_{AR}^2-\Delta_{AR}^2 \omega_{BR}^2)+(-\Delta_{BR}^2 \omega_{AR}^2+\Delta_{AR}^2 \omega_{BR}^2)^2}.
		\end{cases}
		\nonumber
	\end{eqnarray}
We cannot present the results in an analytical form because they are complicated; however,
the solution can be obtained numerically. 
	
Next, by calculating the correlation between the output light $X_{A}(\omega)$ and perturbation of the pendulum mode $\delta \x(\omega)$, we obtain
	\begin{eqnarray}
		S_{X_A \delta \x}(\omega) 
		=
		C_{1X_A \delta \x}\frac{K(\omega)}{|F(\omega)|^2},
	\end{eqnarray}
	where we defined
	\begin{eqnarray}
		K(\omega)
		= 
		(\omega^2
		-\Delta_{BR}^2)  F(\omega)^*+\frac{C_{2X_A\delta \x}}{C_{1X_A\delta \x}} 
		(\omega^2
		-\Delta_{BR}^2)^2,
	\end{eqnarray}
	and
	\begin{align}
		C_{1X_A\delta \x}
		&=
		\sqrt{\eta} \frac{\hbar G_0 \sqrt{n_c} \sqrt{2\kappa}}{M (\kappa^2+\Delta^2)} \left\{ \kappa \left(1-\frac{2 \kappa^2}{\kappa^2+\Delta^2} \right)\langle \A_{\mathrm{in}}^2 (\omega) \rangle -\Delta \frac{2 \kappa \Delta}{\kappa^2+\Delta^2} \langle \B_{\mathrm{in}}^2 (\omega) \rangle \right\}, 
\\
		C_{2X_A\delta \x}
		&=
		- \sqrt{\eta} \frac{2 \hbar^2 G_0^3 (n_c)^{\frac{3}{2}} (2 \kappa)^{\frac{3}{2}} \Delta}{M^2(\kappa^2+\Delta^2)^3}  \left(\kappa^2 \langle \A_{\mathrm{in}}^2 (\omega) \rangle+\Delta^2 \langle \B_{\mathrm{in}}^2 (\omega) \rangle  \right)-\sqrt{\eta}\frac{2 G_0 \sqrt{n_c} \Delta \sqrt{2 \kappa}}{\kappa^2+\Delta^2} \frac{\langle \xi_m^2 (\omega) \rangle}{M^2}.
	\end{align}
The Wiener filter $H_{\delta q}(\omega)$ of the perturbation $\delta \x(\omega)$  is given by
	\begin{align}
		H_{\delta \x}(\omega)
		&=
		\frac{1}{S_{X_AX_A}^+ (\omega)}\left[\frac{S_{X_A \delta \x}(\omega)}{S_{X_A X_A}^-(\omega)} \right]_{+} 
		\nonumber \\
		&=
		\frac{C_{1X_A \delta \x}}{C_{1X_A X_A}} \frac{F(\omega)}{F'(\omega)} \left[\frac{F(\omega)^*}{F'(\omega)^*} \frac{K(\omega)}{F(\omega) F(\omega)^*} \right]_{+} 
		\nonumber \\
		&=
		\frac{C_{1 X_A \delta \x}}{C_{1X_A X_A}} \frac{F(\omega)}{F'(\omega)} \left[\frac{\tilde{E}' \omega^3+\tilde{F}' \omega^2+\tilde{G}' \omega+\tilde{H}'}{F'(\omega)^*}+\frac{\tilde{E} \omega^3+\tilde{F} \omega^2+\tilde{G} \omega+\tilde{H}}{F(\omega)} \right]_{+} 
		\nonumber \\
		&=
		\frac{C_{1 X_A \delta \x}}{C_{1X_A X_A}} \frac{F(\omega)}{F'(\omega)} \frac{\tilde{E} \omega^3+\tilde{F} \omega^2+\tilde{G} \omega+\tilde{H}}{F(\omega)} 
		\nonumber \\
		&=
		\frac{C_{1 X_A \delta \x}}{C_{1X_A X_A}} \frac{\tilde{E} \omega^3+\tilde{F} \omega^2+\tilde{G} \omega+\tilde{H}}{F'(\omega)}
	\end{align}
with the coefficients $\tilde{E},~\tilde{F},~\tilde{G},~\tilde{H},~\tilde{E}',~\tilde{F}',~\tilde{G}',~\tilde{H}'$. We ignored the term proportional to $1/F'(\omega)^*$ in the above derivation because  $[Z(\omega)]_+$ is an operation to take a causal function while the term $1/F'(\omega)^*$ is a non-causal function. The coefficients satisfy
	\begin{eqnarray}
		K(\omega)=(\tilde{E}' \omega^3+\tilde{F}' \omega^2+\tilde{G}' \omega+\tilde{H}')F(\omega)+(\tilde{E} \omega^3+\tilde{F} \omega^2+\tilde{G} \omega+\tilde{H})F'(\omega)^*, 
	\end{eqnarray}
which yields
	\begin{eqnarray}
		\begin{cases}
			{0=-(\tilde{E}+\tilde{E}') ,}
			\\
			{-1=\tilde{A}^* \tilde{E}-\tilde{F}-\tilde{F}'-i \tilde{E}' \gamma_m ,}
			\\
			{i \gamma_m=\tilde{B}^* \tilde{E}+\tilde{A}^* \tilde{F} -\tilde{G}-\tilde{G}'-i \tilde{F}' \gamma_m +\tilde{E}'(\Delta_{BR}^2+\omega_{AR}^2),} 
			\\
			{\frac{C_{2 X_A \delta q}}{C_{1 X_A \delta q}}+2 \Delta_{BR}^2+\omega_{AR}^2}\\
			{\hspace{10pt} =\tilde{C}^* \tilde{E}+\tilde{B}^* \tilde{F}+\tilde{A}^* \tilde{G} -\tilde{H}-\tilde{H}'-i \tilde{G}'\gamma_m +i \tilde{E}' \gamma_m \Delta_{BR}^2+\tilde{F}' (\Delta_{BR}^2+\omega_{AR}^2) ,}
			\\
			{ -2 i \gamma_m \Delta_{BR}^2}\\
			{\hspace{10pt}=\tilde{D}^* \tilde{E}+\tilde{C}^* \tilde{F}+\tilde{B}^* \tilde{G} +\tilde{A}^* \tilde{H}-i \tilde{H}'\gamma_m +i \tilde{F}' \gamma_m \Delta_{BR}^2+\tilde{G}' (\Delta_{BR}^2+\omega_{AR}^2)}\\
			{\hspace{20pt}+\tilde{E}' (-\Delta_{BR}^2 \omega_{AR}^2+\Delta_{AR}^2 \omega_{BR}^2),}
			\\
			{-2 \frac{C_{2 X_A \delta q}}{C_{1 X_A \delta q}} \Delta_{BR}^2-\Delta_{BR}^4 -2 \Delta_{BR}^2 \omega_{AR}^2+ \Delta_{AR} ^2 \omega_{BR}^2}\\
			{\hspace{10pt}=\tilde{D}^* \tilde{F}+\tilde{C}^* \tilde{G}+\tilde{B}^* \tilde{H}  +i \tilde{G}' \gamma_m \Delta_{BR}^2+\tilde{H}' (\Delta_{BR}^2+\omega_{AR}^2)+\tilde{F}' (-\Delta_{BR}^2 \omega_{AR}^2+\Delta_{AR}^2 \omega_{BR}^2)},
			\\
			{i \gamma_m \Delta_{BR}^4 =\tilde{D}^* \tilde{G}+\tilde{C}^* \tilde{H} +i \tilde{H}' \gamma_m \Delta_{BR}^2+\tilde{G}' (-\Delta_{BR}^2 \omega_{AR}^2+\Delta_{AR}^2 \omega_{BR}^2)},
			\\
			{\frac{C_{2 X_A \delta q}}{C_{1 X_A \delta q}} \Delta_{BR}^4+\Delta_{BR}^4 \omega_{AR}^2 -\Delta_{AR}^2 \Delta_{BR}^2 \omega_{BR}^2=\tilde{D}^*\tilde{H}+\tilde{H}'(- \Delta_{BR}^2 \omega_{AR}^2+ \Delta_{AR}^2 \omega_{BR}^2)}.
		\end{cases}
		\nonumber
	\end{eqnarray}
The coefficients $\tilde{E},~\tilde{F},~\tilde{G},~\tilde{H},~\tilde{E}',~\tilde{F}',~\tilde{G}',~\tilde{H}'$ can be obtained by solving the above equations. However, the explicit expressions are quite complicated, and we find the solution numerically.
	
The Wiener filter $H_{\delta p}(\omega)$ for momentum $\delta p(\omega)$ can be obtained in the same manner because $\delta p(\omega)=-i M \omega \delta \x(\omega)$
	\begin{align}
		S_{X_A \delta p}(\omega)
		&=
		-i M \omega S_{X_A \delta \x}(\omega), 
		\\
		H_{\delta p}(\omega)
		&=
		\frac{1}{S_{X_AX_A}^+ (\omega)}\left[\frac{S_{X_A \delta p}(\omega)}{S_{X_A X_A}^-(\omega)} \right]_{+} 
		\nonumber \\
		&=
		\frac{C_{1 X_A \delta \x}}{C_{1X_A X_A}} \frac{F(\omega)}{F'(\omega)} \left[\frac{F(\omega)^*}{F'(\omega)^*} \frac{-i M \omega K(\omega)}{F(\omega) F(\omega)^*} \right]_{+} 
		\nonumber \\
		&=
		\frac{C_{1 X_A \delta \x}}{C_{1X_A X_A}} \frac{F(\omega)}{F'(\omega)} \left[\frac{\tilde{I}' \omega^3+\tilde{J}' \omega^2+\tilde{K}' \omega+\tilde{L}'}{F'(\omega)^*}+\frac{\tilde{I} \omega^3+\tilde{J} \omega^2+\tilde{K} \omega+\tilde{L}}{F(\omega)} \right]_{+} 
		\nonumber \\
		&=
		\frac{C_{1 X_A \delta \x}}{C_{1X_A X_A}} \frac{F(\omega)}{F'(\omega)} \frac{\tilde{I} \omega^3+\tilde{J} \omega^2+\tilde{K} \omega+\tilde{L}}{F(\omega)} 
		\nonumber \\
		&=
		\frac{C_{1 X_A \delta \x}}{C_{1X_A X_A}} \frac{\tilde{I} \omega^3+\tilde{J} \omega^2+\tilde{K} \omega+\tilde{L}}{F'(\omega)}
	\end{align}
		with coefficients $\tilde{I},~\tilde{J},~\tilde{K},~\tilde{L}$, which 
		are obtained numerically, similar to that described above.

Then, by estimating the correlation between the output light $X_{A}$ and perturbation of the mirror rotation $\delta \Phi(\omega)$, we obtain:
	\begin{eqnarray}
		S_{X_A \delta \Phi}(\omega)
		=
		C_{1X_A\delta \x} \frac{L(\omega)}{|F(\omega)|^2},
	\end{eqnarray}
	where we defined
	\begin{eqnarray}
		L(\omega)
		=
		-\omega_B^2   F(\omega)^*-\omega_{BR}^2 \frac{C_{2X_A\delta \x}}{C_{1X_A\delta \x}} (\omega^2
		-\Delta_{BR}^{2}).
	\end{eqnarray}
	The Wiener filter $H_{\delta \Phi}(\omega)$ for the perturbation  of the mirror rotation $\delta \Phi(\omega)$ is given by
	\begin{align}
		H_{\delta \Phi}(\omega)
		&=
		\frac{1}{S_{X_AX_A}^+ (\omega)}\left[\frac{S_{X_A \delta \Phi}(\omega)}{S_{X_A X_A}^-(\omega)} \right]_{+} 
		\nonumber \\
		&=
		\frac{C_{1X_A\delta \x}}{C_{1X_A X_A}} \frac{F(\omega)}{F'(\omega)} \left[\frac{F(\omega)^*}{F'(\omega)^*} \frac{L(\omega)}{F(\omega) F(\omega)^*} \right]_{+} 
		\nonumber \\
		&=
		\frac{C_{1 X_A \delta \x}}{C_{1X_A X_A}} \frac{F(\omega)}{F'(\omega)} \left[\frac{\tilde{M}' \omega^3+\tilde{N}' \omega^2+\tilde{O}' \omega+\tilde{P}'}{F'(\omega)^*}+\frac{\tilde{M} \omega^3+\tilde{N} \omega^2+\tilde{O} \omega+\tilde{P}}{F(\omega)} \right]_{+} 
		\nonumber \\
		&=
		\frac{C_{1 X_A \delta \x}}{C_{1X_A X_A}} \frac{F(\omega)}{F'(\omega)} \frac{\tilde{M} \omega^3+\tilde{N} \omega^2+\tilde{O} \omega+\tilde{P}}{F(\omega)} 
		\nonumber \\
		&=
		\frac{C_{1 X_A \delta \x}}{C_{1X_A X_A}} \frac{\tilde{M} \omega^3+\tilde{N} \omega^2+\tilde{O} \omega+\tilde{P}}{F'(\omega)},
	\end{align}
		where $\tilde{M},~\tilde{N},~\tilde{O},~\tilde{P}$  
		are numerically obtained, similar to those described above.
	
	The Wiener filter for the momentum perturbation conjugate to the mirror rotation $\delta \Pi_{\Phi}(\omega)$ can be obtained similarly because $\delta \Pi_{\Phi}(\omega)=- i J \omega \delta \Phi(\omega)$,
	\begin{align}
		S_{X_A \delta \Pi_{\Phi}}(\omega)
		&=
		-i J \omega S_{X_A \delta \Phi}(\omega), 
		\\
		H_{\delta \Pi_{\Phi}}(\omega)
		&=
		\frac{1}{S_{X_AX_A}^+ (\omega)}\left[\frac{S_{X_A \delta \Pi_{\Phi}}(\omega)}{S_{X_A X_A}^-(\omega)} \right]_{+} 
		\nonumber \\
		&=
		\frac{C_{1X_A\delta \x}}{C_{1X_A X_A}} \frac{F(\omega)}{F'(\omega)} \left[\frac{F(\omega)^*}{F'(\omega)^*} \frac{-iJ L(\omega)}{F(\omega) F(\omega)^*} \right]_{+} 
		\nonumber \\
		&=
		\frac{C_{1 X_A \delta \x}}{C_{1X_A X_A}} \frac{F(\omega)}{F'(\omega)} \left[\frac{\tilde{Q}' \omega^3+\tilde{R}' \omega^2+\tilde{S}' \omega+\tilde{T}'}{F'(\omega)^*}+\frac{\tilde{Q} \omega^3+\tilde{R} \omega^2+\tilde{S} \omega+\tilde{T}}{F(\omega)} \right]_{+} 
		\nonumber \\
		&=
		\frac{C_{1 X_A \delta \x}}{C_{1X_A X_A}} \frac{F(\omega)}{F'(\omega)} \frac{\tilde{Q} \omega^3+\tilde{R} \omega^2+\tilde{S} \omega+\tilde{T}}{F(\omega)} 
		\nonumber \\
		&=
		\frac{C_{1 X_A \delta \x}}{C_{1X_A X_A}} \frac{\tilde{Q} \omega^3+\tilde{R} \omega^2+\tilde{S} \omega+\tilde{T}}{F'(\omega)},
	\end{align}
	{
		where the coefficients $\tilde{Q},~\tilde{R},~\tilde{S},~\tilde{T}$ 
		are obtained numerically.
	}

\section{{
Wiener filter for the point-particle mirror}\label{point}}
We here review the Wiener filter presented in a previous study~\cite{MY} and compare the results with those of our beam model. It is noteworthy that the mirror in Ref.~\cite{MY} was treated as a point particle, in contrast to the rigid body used in the present study.
The formulation in this appendix is essentially identical to that described in Refs.~\cite{MY}, but
we present a formulation using dimensional quantities, which is useful for 
comparing with the results of the two-mode theory. 
	
	\subsection{Model and Systems}
 \label{pointone}
We consider the Hamiltonian of a system consisting of a mirror and cavity photons described by 
\begin{align}
H=\frac{1}{2M} p_m^2+\frac{1}{2} M \Omega^2 q_m^2+\frac{\hbar \Delta}{2}(x_m^2+y_m^2)- \hbar G_0 \sqrt{n_c} x_m q_m,
\end{align}
where $q_m$ and $p_m$ are the positions and canonical momentum, respectively, of the oscillator of the mirror with angular frequency $\Omega$. $x_m$ and $y_m$ are the amplitude and phase quadratures of the cavity photons, respectively.
Following the Heisenberg equation, the equations of motion can be derived as follows. 
Using the perturbative method and setting 
$\x_m=\bar{\x}_m+\delta \x_m(t)$, $p_m=\delta p_m(t)$, $x_m=\bar{x}_m+\delta x_m(t)$, and 
$y_m=\bar{y}_m+\delta y_m(t)$, we can write the Langevin equations for the 
perturbative quantities as 
	\begin{align}
		\delta \dot{q}_{\text{m}}
		&=
		\frac{1}{M} \delta p_m, 
		\\
		\delta \dot{p}_{\text{m}}
		&=
		-M \Omega^2 \delta q_{\text{m}}-\Gamma \delta p_{\text{m}}+\xi+\hbar G_0 \sqrt{n_c} \delta x_{\text{m}} -\int_{-\infty}^{\infty} ds g_{\mathrm{FB}}(t-s) X_{\text{m}}(s) \sqrt{\frac{\hbar M \Omega}{2}}, 
		\label{p_matsu} 
		\\
		\delta \dot{x}_{\text{m}}
		&=
		-\kappa \delta x_{\text{m}}+\Delta \delta y_{\text{m}}+\sqrt{2 \kappa}x_{\mathrm{in}}, \label{x_matsu} 
		\\
		\delta \dot{y}_{\text{m}}
		&=
		-\kappa \delta y_{\text{m}}-\Delta \delta x_{\text{m}}+\sqrt{2 \kappa} y_{\mathrm{in}}+2 G_0 \sqrt{n_c} \delta q_{\text{m}}, 
		\label{y_matsu} 
	\end{align}
where the noises $\xi$, $x_{\rm in}$ and $y_{in}$ are included with variance,
where $\langle x_{\text{in}}^2 \rangle=\langle y_{\text{in}}^2 \rangle=2 N(\omega_c)+1$ and $\langle \xi^2 \rangle \simeq \hbar M \Omega \Gamma  \langle p_{\mathrm{in}}^2 \rangle$ with $\langle p_{\text{in}}^2 \rangle=2 k_\text{B} T_0/\hbar \Omega +1$.
The last term of Eq.~(\ref{p_matsu}) represents the feedback cooling.
The output light, denoted $X_{\text{m}}(t)$, satisfies 
the following input-output relations:
	\begin{align}
		X_{\text{m}} 
		=\sqrt{\eta} x_{\mathrm{m, out}}+\sqrt{1-\eta} x_{\mathrm{in}}^{'},
		\quad
		x_{\mathrm{m, out}}
		=
		x_{\mathrm{in}}-\sqrt{2 \kappa} 
		\delta x_{\text{m}},
	\end{align}
	where $\eta \in [0,1]$ is the detection efficiency and $x^{'}_{\text{in}}$ is introduced owing to imperfect detection.
	Assuming the adiabatic approximation ($\kappa \gg \omega$),
	Eqs.~(\ref{x_matsu}) and~(\ref{y_matsu}) can be solved as
	\begin{align}
		\delta x_m
		&= 
		\frac{\sqrt{2 \kappa}}{\kappa^2+\Delta^2}(\kappa x_{\mathrm{in}}+\Delta y_{\mathrm{in}})+\frac{2 \Delta G_0 \sqrt{n_c}}{\kappa^2+\Delta^2} \delta q_m,
		\quad
		\delta y_m
		=\frac{\sqrt{2 \kappa}}{\kappa^2+\Delta^2} (\kappa y_{\mathrm{in}}-\Delta x_{\mathrm{in}})+\frac{2 \kappa G_0 \sqrt{n_c}}{\kappa^2+\Delta^2}\delta q_m.
	\end{align}
	Using the above equations, the Langevin equations can be summarized as 
	\begin{align}
		\delta \dot{q}_{\text{m}}
		&=\frac{1}{M} \delta p_{\text{m}},
		\\
		\delta \dot{p}_{\text{m}}
		&=
		-M\left( \Omega^2 -\frac{2 \hbar G_0^2 n_c \Delta }{M(\kappa^2+\Delta^2)} \right) \delta q_{\text{m}}-\gamma_m \delta p_m+ \xi_{\text{m}}+\frac{\hbar G_0 \sqrt{2\kappa} \sqrt{n_\text{c}}}{\kappa^2+\Delta^2} (\kappa x_{\mathrm{in}}+\Delta y_{\mathrm{in}}), 
		\\
		X_{\text{m}}
		&=
		\sqrt{\eta} \left\{\left(1-\frac{2 \kappa^2}{\kappa^2+\Delta^2}\right)x _{\mathrm{in}}-\frac{2 \kappa \Delta}{\kappa^2+\Delta^2} y_{\mathrm{in}}-\frac{2 \Delta G_0 \sqrt{2 \kappa} \sqrt{n_\text{c}}}{\kappa^2+\Delta^2}\delta q_{\text{m}} \right\}+\sqrt{1-\eta} x_{\mathrm{in}}^{'}.
	\end{align}
	Here, the feedback cooling term was replaced by the dissipation $\gamma_m$, and the fluctuation force $\xi$ was changed to $\xi_\text{m}$ with $\langle \xi^2_\text{m} \rangle=\hbar M \Omega \gamma_\text{m} \langle p_{\text{in}}'^2 \rangle$ and $\langle p_{\text{in}}'^2 \rangle=2 k_\text{B} T_0 \Gamma/\hbar \Omega \gamma_\text{m}+1$.
	
	\subsection{Spectral density and Wiener filter}
 \label{onemodeWF}
	{
		We solved the above equations in the Fourier space and calculated the spectral density to obtain the Wiener filter. The Langevin equations in the Fourier space are solved as}
	\begin{align}
		\delta q_\text{m}(\omega)
		&=
		\frac{1}{M F_m(\omega)} \left\{\xi_m(\omega)
		+\frac{
			\hbar G_0 \sqrt{n_c}\sqrt{2\kappa}}{\kappa^2+\Delta^2}(\kappa x_{\mathrm{in}}(\omega)+\Delta y_{\mathrm{in}}(\omega)) \right\}, 
		\\
		\delta p_\text{m}(\omega)
		&=
		-i \omega M \delta q_\text{m}(\omega)
		= -\frac{i \omega}{ F_m(\omega)} \left\{\xi_m(\omega)
		+\frac{
			\hbar G_0 \sqrt{n_c}\sqrt{2\kappa}}{\kappa^2+\Delta^2}(\kappa x_{\mathrm{in}}(\omega)+\Delta y_{\mathrm{in}}(\omega)) \right\}, 
		\\
		X_\text{m}(\omega)
		=
		\sqrt{\eta} 
		\Big[
		&
		-\left(\frac{\kappa^2-\Delta^2 }{\kappa^2+\Delta^2}+\frac{4 \kappa^2 \Delta \hbar G_0^2 n_c }{M F_\text{m}(\omega)(\kappa^2+\Delta^2)^2}\right)x_{\mathrm{in}}(\omega) 
		-\left(\frac{2 \kappa \Delta}{\kappa^2+\Delta^2}+\frac{4 \kappa \Delta^2 \hbar G_0^2 n_c}{M F_\text{m}(\omega)(\kappa^2+\Delta^2)^2}
		\right)y_{\mathrm{in}}(\omega) 
		\nonumber \\
		&
		-\frac{2 \Delta G_0 \sqrt{2 \kappa} \sqrt{n_c}}{MF_\text{m}(\omega)(\kappa^2+\Delta^2)} \xi_\text{m}(\omega)
		\Big]
		+\sqrt{1-\eta} x_{\mathrm{in}}^{'}(\omega).
	\end{align}
	We defined $F_{\text{m}}(\omega)$ and $\omega_{\text{m}}$ as
	\begin{eqnarray}
		F_{\text{m}}(\omega)=\omega_m^2 -i \gamma_m \omega -\omega^2,
		\quad
		\omega_{\text{m}}=\sqrt{\Omega^2-\frac{2 \hbar G_0^2 n_c \Delta}{M(\kappa^2+\Delta^2)} }.
	\end{eqnarray}
	The spectral density $S_{X_\text{m} X_\text{m}}(\omega)$ with respect to $X_\text{m}(\omega)$ is:
	\begin{eqnarray}
		S_{X_\text{m} X_\text{m}}(\omega)
		=
		C_{1 X_\text{A} X_\text{A}} \frac{J_\text{m}(\omega)}{|F_\text{m}(\omega)|^2},
	\end{eqnarray}
	where we defined
	\begin{eqnarray}
		J_\text{m}(\omega)
		=
		|F_\text{m}(\omega)|^2+\frac{C_{2 X_\text{A} X_\text{A}}}{C_{1 X_\text{A} X_\text{A}}}(F_\text{m}(\omega)+F^*_\text{m}(\omega))+\frac{C_{3 X_\text{A} X_\text{A}}}{C_{1 X_\text{A} X_\text{A}}},
	\end{eqnarray}
	$J_\text{m} (\omega)=\omega^4+\alpha \omega^2+\beta$, and the coefficients $\alpha$ and $\beta$ are given by
	\begin{eqnarray}
		\alpha=
		-2 \frac{C_{2 X_\text{A} X_\text{A}}}{C_{1 X_\text{A} X_\text{A}}}+\gamma^2_\text{m} -2 \omega^2_\text{m},
		\quad
		\beta=\frac{C_{3 X_\text{A} X_\text{A}}}{C_{1 X_\text{A} X_\text{A}}}+2\frac{C_{2 X_\text{A} X_\text{A}}}{C_{1 X_\text{A} X_\text{A}}} \omega^2_\text{m}+\omega^4_\text{m}.
	\end{eqnarray}
	The spectral density $S_{X_\text{m} X_\text{m}}(\omega)$ consists of a causal $S_{X_\text{m} X_\text{m}}^+(\omega)$
        and non-causal $S_{X_\text{m} X_\text{m}}^-(\omega)$ part.
	\begin{eqnarray}
		S_{X_\text{m} X_\text{m}}(\omega) 
		=
		S_{X_\text{m} X_\text{m}}^{+}(\omega) S_{X_\text{m} X_\text{m}}^{-}(\omega),
		\quad
		S_{X_\text{m} X_\text{m}}^{+}(\omega)
		=
		\sqrt{C_{1 X_\text{A} X_\text{A}}} \frac{F'_\text{m}(\omega)}{F_\text{m}(\omega)},
		\quad
		S_{X_\text{m} X_\text{m}}^{-}(\omega)
		=
		\sqrt{C_{1 X_\text{A} X_\text{A}}} \frac{F'_\text{m}(\omega)^*}{F_\text{m}(\omega)^*} \label{sd_Xm},
	\end{eqnarray}
	where $F'_\text{m}(\omega)$ is a causal function, expressed as follows:
	\begin{eqnarray}
		F'_\text{m}(\omega)=\Omega'^2 { - } i\Gamma' \omega -\omega^2
	\end{eqnarray}
with coefficients $\Gamma'=\sqrt{\alpha+2\sqrt{\beta}}$ and $\Omega'^2=\sqrt{\beta}$.
The correlation between the output light $X_{\text{m}}(\omega)$ and center of mass position $\delta q_{\text{m}}(\omega)$ is expressed as
	\begin{align}
		S_{X_\text{m} \delta q_\text{m}}(\omega)
		&=
		C_{1 X_\text{A} \delta \x} \frac{K_\text{m}(\omega)}{|F_\text{m}(\omega)|^2}, 
		\\
		K_\text{m}(\omega)
		&=
		F^*_\text{m}(\omega)+\frac{C_{2 X_\text{A} \delta \x}}{C_{1 X_\text{A} \delta \x}}
		\equiv
		\tilde{\Omega}^2- i \gamma_\text{m} {\omega}-\omega^2, 
		\quad
		\tilde{\Omega}^2
		=
		\omega_\text{m}^2+\frac{C_{2 X_\text{A} \delta \x}}{C_{1 X_\text{A} \delta \x}}.
	\end{align}
	Subsequently, the Wiener filter with respect to $\delta q_{\text{m}}(\omega)$ is calculated as follows:
	\begin{align}
		H_{\delta q_\text{m}}(\omega)
		&=
		\frac{1}{S_{X_\text{m} X_\text{m}}^+ (\omega)}\left[\frac{S_{X_\text{m} \delta q_\text{m}}(\omega)}{S_{X_\text{m} X_\text{m}}^-(\omega)} \right]_{+} 
		\nonumber \\
		&=
		\frac{C_{1X_\text{A} \delta \x}}{C_{1X_\text{A} X_\text{A}}} \frac{F_\text{m}(\omega)}{F'_\text{m}(\omega)} \left[\frac{F_\text{m}(\omega)^*}{F'_\text{m}(\omega)^*} \frac{K_\text{m}(\omega)}{F_\text{m}(\omega) F_\text{m}(\omega)^*} \right]_{+} 
		\nonumber \\
		&=
		\frac{C_{1 X_\text{A} \delta \x}}{C_{1X_\text{A} X_\text{A}}} \frac{F_\text{m}(\omega)}{F'_\text{m}(\omega)} \left[\frac{\tilde{E}'_\text{m} \omega+\tilde{F}'_\text{m}}{F'_\text{m}(\omega)}+\frac{\tilde{E}_\text{m} \omega+\tilde{F}_\text{m}}{F_\text{m}(\omega)} \right]_{+} 
		\nonumber \\
		&=
		\frac{C_{1 X_\text{A} \delta \x}}{C_{1X_\text{A} X_\text{A}}} \frac{F_\text{m}(\omega)}{F'_\text{m}(\omega)} \frac{\tilde{E}_\text{m} \omega+\tilde{F}_\text{m}}{F_\text{m}(\omega)} 
		\nonumber \\
		&=
		\frac{C_{1 X_\text{A} \delta \x}}{C_{1X_\text{A} X_\text{A}}} \frac{\tilde{E}_\text{m} \omega+\tilde{F}_\text{m}}{F'_\text{m}(\omega)}.
	\end{align}
	Here, the coefficients $\tilde{E}_\text{m},\tilde{F}_\text{m}$ are expressed as
	\begin{eqnarray}
		\tilde{E}_\text{m}
		=
		i \frac{(\gamma_\text{m}+\Gamma')(\omega^2_\text{m}-\tilde{\Omega}^2)}{(\Gamma' \omega^2_\text{m}+\gamma_\text{m} \Omega'^2)(\gamma_\text{m}+\Gamma')+(\omega^2_\text{m}-\Omega'^2)^2},
		\quad
		\tilde{F}_\text{m}
		=
		\frac{(\omega^2_\text{m}-\gamma^2_\text{m}-\gamma_\text{m} \Gamma'-\Omega'^2)(\omega^2_\text{m}-\tilde{\Omega}^2)}{(\Gamma' \omega^2_\text{m}+\gamma_\text{m} \Omega'^2)(\gamma_\text{m}+\Gamma')+(\omega^2_\text{m}-\Omega'^2)^2}.
	\end{eqnarray}
	Because $\delta p_\text{m}(\omega)=- i M \omega \delta q_\text{m}(\omega)$, the Wiener filter with respect to $\delta p_\text{m}$ can be similarly obtained.
	\begin{align}
		S_{X_\text{m} \delta p_\text{m}}(\omega)
		&=
		-i M \omega S_{X_\text{m} \delta q_\text{m}}(\omega), 
		\\
		H_{\delta p_\text{m}}
		&=
		\frac{1}{S_{X_\text{m} X_\text{m}}^{+} (\omega)}\left[\frac{S_{X_\text{m} \delta p_\text{m}}(\omega)}{S_{X_\text{m} X_\text{m}}^-(\omega)} \right]_{+} 
		\nonumber \\
		&=
		\frac{C_{1 X_\text{A} \delta \x}}{C_{1X_\text{A} X_\text{A}}} \frac{F_\text{m}(\omega)}{F'_\text{m}(\omega)} \left[\frac{F_\text{m}(\omega)^*}{F'_\text{m}(\omega)^*} \frac{-i M \omega K_\text{m}(\omega)}{F_\text{m}(\omega) F_\text{m}(\omega)^*} \right]_{+} 
		\nonumber \\
		&=
		\frac{C_{1 X_\text{A} \delta \x}}{C_{1X_\text{A} X_\text{A}}} \frac{F_\text{m}(\omega)}{F'_\text{m}(\omega)} \left[\frac{\tilde{I}'_\text{m} \omega+\tilde{J}'_\text{m}}{F'_\text{m}(\omega)}+\frac{\tilde{I}_\text{m} \omega+\tilde{J}_\text{m}}{F_\text{m}(\omega)} \right]_{+} 
		\nonumber \\
		&=
		\frac{C_{1 X_\text{A} \delta \x}}{C_{1X_\text{A} X_\text{A}}} \frac{F_\text{m}(\omega)}{F'_\text{m}(\omega)} \frac{\tilde{I}_\text{m} \omega+\tilde{J}_\text{m}}{F_\text{m}(\omega)}
		\nonumber \\
		&=
		\frac{C_{1 X_\text{A} \delta \x}}{C_{1X_\text{A} X_\text{A}}} \frac{\tilde{I}_\text{m} \omega+\tilde{J}_\text{m}}{F'_\text{m}(\omega)},
	\end{align}
	where the coefficients $\tilde{I}_\text{m},\tilde{J}_\text{m}$ are  given by
	\begin{eqnarray}
		\tilde{I}_\text{m}
		=
		-i M \omega_m^2 \tilde{E}_{\text{m}}, 
		\quad
		\tilde{J}_\text{m}
		=
		M \frac{\Omega'^2-\omega^2_\text{m}}{\Gamma'+\gamma_\text{m}} \tilde{E}_\text{m}.
	\end{eqnarray}
	
	\subsection{One-mode conditional covariance}
	The causal part of the spectral density of the output light $S_{X_\text{m} X_\text{m}}^{+}(\omega)$, obtained using the one-mode model, is given by Eq.~(\ref{sd_Xm}), and the following $G_{\delta q_m}(\omega)$ and $G_{\delta p_m}(\omega)$ can be calculated by applying filters $H_{\delta q_\text{m}}(\omega), H_{\delta p_\text{m}}(\omega)$ when the mirror is treated as a mass point
	\begin{align}
		G_{\delta q_m}(\omega)
		&=
		H_{\delta q_\text{m}}(\omega) S_{X_\text{A} X_\text{A}}^{+}(\omega)
		=\frac{C_{1 X_\text{A} \delta \x}}{\sqrt{C_{1X_\text{A} X_\text{A}}}} 
		\frac{F'(\omega)}{F(\omega)}
		\frac{\tilde{E}_\text{m} \omega+\tilde{F}_\text{m}}{F'_\text{m}(\omega)}, 
\\
		\quad
		G_{\delta p_m}(\omega)
		&=
		H_{\delta p_\text{m}}(\omega) S_{X_\text{A} X_\text{A}}^{+}(\omega)
		=\frac{C_{1 X_\text{A} \delta \x}}{\sqrt{C_{1X_\text{A} X_\text{A}}}} 
		\frac{F'(\omega)}{F(\omega)}
		\frac{\tilde{I}_\text{m} \omega+\tilde{J}_\text{m}}{F'_\text{m}(\omega)}. 
	\end{align}
	From this, the conditional variance matrix of the Wiener filter when the mirror is treated as a mass point is given by
	\begin{eqnarray}
		\bm{V}_{\text{cm}}
		=
		\begin{pmatrix}
			V_{\delta q_m \delta q_m} & V_{\delta q_m 
				\delta p_m} 
			\\ 
			V_{\delta q_m \delta p_m} & V_{\delta p_m \delta p_m}
			\label{Vcm}
		\end{pmatrix},
	\end{eqnarray}
	where the components $V_{\delta q_m \delta q_m}$, $V_{\delta q_m \delta p_m}$, and $V_{\delta p_m \delta p_m}$ are obtained by
	\begin{align}
		V_{\delta q_m \delta q_m}
		&=
		\frac{1}{2\pi} \int_{-\infty}^{\infty} \mathrm{Re}[S_{\delta q_m \delta q_m}(\omega) -|G_{\delta q_m}(\omega)|^2)]d \omega, 
		\label{Vcmqq}
		\\
		V_{\delta p_m \delta p_m}&=
		\frac{1}{2\pi} \int_{-\infty}^{\infty} \mathrm{Re}[S_{\delta p_m \delta p_m}(\omega) -|G_{\delta p_m}(\omega)|^2] d \omega, 
		\label{Vcmpp}
		\\
		V_{\delta q_m \delta p_m}
		&=
		\frac{1}{2\pi} \int_{-\infty}^{\infty} \mathrm{Re} [S_{\delta q_m \delta p_m}(\omega) -G_{\delta q_m}(\omega)^* G_{\delta p_m}(\omega)] d \omega.
		\label{Vcmqp}
	\end{align}
	Here, we use the following equations:
	\begin{align}
		S_{\delta q_m \delta q_m}(\omega)
		&=\frac{1}{M^2 |F_m(\omega)|^2} \left\{ \langle\xi_m^2(\omega) \rangle+\frac{2 \hbar^2 G_0^2 n_c \kappa}{\kappa^2+\Delta^2}(\kappa^2 \langle x_{\text{in}}^2(\omega) \rangle+\Delta^2 \langle y_{\text{in}}^2(\omega) \rangle) \right\}, 
		\\
		S_{\delta p_m \delta p_m}(\omega)
		&= M^2 \omega^2 S_{\delta q_m \delta q_m}(\omega) , 
		\\  
		S_{\delta x_m \delta p_m}(\omega)
		&=i M \omega S_{\delta q_m \delta q_m}(\omega).
	\end{align}
		The conditional covariance $V_{\mathrm{cm}}$ is the same as that developed in Ref.~\cite{MY}.
		
\subsection{Conditional covariance with the one-mode Wiener filter for the two-mode model} 
\label{Appenfour}
Here we demonstrate a case where the system is imperfectly modeled.  
Let us consider the case in which the other Wiener filters $H_{\delta q_m}(\omega), H_{\delta p_m}(\omega)$ based on the one-mode theory of only the pendulum mode, is adopted for the light $X_A(\omega)$ obtained in the beam model.
		In this case, the other Wiener filters $H_{\delta q_m}(\omega)$ and $H_{\delta p_m}(\omega)$ are not optimal estimates. Therefore, returning to the expression of the squared mean difference between the true and estimated values, we obtain
		\begin{eqnarray}
			\langle|\delta q(\omega)-\delta q_{\text{em}}(\omega)|^2\rangle
			&=&S_{\delta q \delta q}(\omega)-H_{\delta q_m}^{*}(\omega) S_{X_A \delta q}(\omega)-H_{\delta q_m}(\omega) S_{\delta q X_A}(\omega)+|H_{\delta q_m}(\omega)|^2 S_{X_A X_A}(\omega),
		\end{eqnarray}
		where we use
		\begin{eqnarray}
			\delta q_{\text{em}}(\omega) &\equiv& H_{\delta q_m}(\omega) X_A(\omega)  \\
			S_{X_A \delta q}(\omega)&=& S_{\delta q X_A}^*(\omega). 
		\end{eqnarray}
		
Similarly, for the perturbation of the momentum of the center of mass of the mirror $\delta p(\omega)$, the mean-squared difference between the true and estimated values $\langle|\delta p(\omega)-\delta p_{\rm em}(\omega)|^2\rangle$ is
		\begin{eqnarray}
			\langle|\delta p(\omega)-\delta p_{\text{em}}(\omega)|^2\rangle
			&=&S_{\delta p \delta p}(\omega)-H_{\delta p_m}^{*}(\omega) S_{X_A \delta p}(\omega)-H_{\delta p_m}(\omega) S_{\delta p X_A}(\omega)+|H_{\delta p_m}(\omega)|^2 S_{X_A X_A}(\omega),
		\end{eqnarray}
		where we use
		\begin{eqnarray}
			\delta p_{\text{em}}(\omega) &\equiv& H_{\delta p_m}(\omega) X_A(\omega) 
   \\
			S_{X_A \delta p}(\omega)&=& S_{\delta p X_A}^*(\omega).
		\end{eqnarray}
		
For the correlation between the perturbation of the  center of gravity position of the mirror $\delta q(\omega)$ and the perturbation of its momentum $\delta p(\omega)$, the correlation $\langle(\delta q(\omega)-\delta q_{\rm em}(\omega))^{\dag}(\delta p(\omega)-\delta p_{\rm em}(\omega))+(\delta p(\omega)-\delta p_{\rm em}(\omega))(\delta q(\omega)-\delta q_{\rm em}(\omega))^{\dag}\rangle$ is written as follows:
		\begin{eqnarray}
			&&\langle (\delta q(\omega)-\delta q_{\text{em}}(\omega))^{\dag} (\delta p(\omega)-\delta p_{\text{em}}(\omega))+(\delta p(\omega)-\delta p_{\rm em}(\omega))(\delta q(\omega)-\delta q_{\rm em}(\omega))^{\dag} \rangle \nonumber \\
			&&~~~~~=S_{\delta q \delta p}(\omega)-H_{\delta q_m}^{*}(\omega) S_{X_A \delta p}(\omega)-H_{\delta p_m}(\omega) S_{\delta q X_A}(\omega)+H_{\delta q_m}^*(\omega) H_{\delta p_m}(\omega) S_{X_A X_A}(\omega).
		\end{eqnarray}
		These are inverse Fourier-transformed into real space to obtain the conditional covariance $\bm{V}'_{\text{c}}$, which is expressed as follows:
		\begin{eqnarray}
		\bm{V}'_{\text{c}}=
		\begin{pmatrix}
				V'_{\delta q \delta q} & V'_{\delta q \delta p} \\
				V'_{\delta q \delta p} & V'_{\delta p \delta p} \\
			\end{pmatrix}, \label{V'_c}
		\end{eqnarray}
		where
\begin{eqnarray}
V'_{\delta q \delta q}&=&\frac{1}{2\pi} \int_{-\infty}^{\infty} d\omega \mathrm{Re}[S_{\delta q \delta q}(\omega)-H_{\delta q_m}^{*}(\omega) S_{X_A \delta q}(\omega)-H_{\delta q_m}(\omega) S_{\delta q X_A}(\omega)+|H_{\delta q_m}(\omega)|^2 S_{X_A X_A}(\omega)], 
\label{V'_dqdq} \\
V'_{\delta p \delta p}&=&\frac{1}{2\pi} \int_{-\infty}^{\infty} d\omega \mathrm{Re}[S_{\delta p \delta p}(\omega)-H_{\delta p_m}^{*}(\omega) S_{X_A \delta p}(\omega)-H_{\delta p_m}(\omega) S_{\delta p X_A}(\omega)+|H_{\delta p_m}(\omega)|^2 S_{X_A X_A}(\omega)], 
\label{V'_dpdp} \\
V'_{\delta q \delta p}&=&\frac{1}{2\pi} \int_{-\infty}^{\infty} d\omega \mathrm{Re}[S_{\delta q \delta p}(\omega)-H_{\delta q_m}^{*}(\omega) S_{X_A \delta p}(\omega)-H_{\delta p_m}(\omega) S_{\delta q X_A}(\omega) 
+H_{\delta q_m}^*(\omega) H_{\delta p_m}(\omega) S_{X_A X_A}(\omega) ]. \nonumber \\ \label{V'_dqdp} 
\end{eqnarray}


\begin{thebibliography}{}
			\bibitem{Aspelmeyer}
			M. Aspelmeyer, T. J. Kippenberg, and F. Marquardt,
			Rev. Mod. Phys. ${\bf 86 }$, 1391 (2014).
			
			\bibitem{Bowen}
			W. P. Bowen and G. J. Milburn, Quantum Optomechanics (CRC Press, 2016).
			
			\bibitem{Yambei}
			Y. Chen,
			J. Phys. B: At. Mol. Opt. Phys. {\bf 46} 104001 (2013).
			
			\bibitem{Michimura}
			Y. Michimura and K. Komori,
			Eur. Phys. J. D {\bf 74} 6, 126 (2020).
			\bibitem{Croquette}
            M. Croquette, et al.,  AVS Quantum Science, {\bf 5}, 014403 (2023)
            
			\bibitem{Korppi}
			C. F. Ockeloen-Korppi, E. Damsk\"{a}gg, J. -M. Pirkkalainen, M. Asjad, A. A. Clerk, F. Massel, M. J. Woolley, and M. A. Sillanp\"{a}\"{a},
			Stabilized entanglement of massive mechanical oscillators
			Nature {\bf 556}, 478 (2018).
			\bibitem{Kotler}
			S. Kotler, G. A. Peterson, E. Shojaee, F. Lecocq, K. Cicak, A. Kwiatkowski, S. Geller, S. Glancy, E. Knill, R. W. Simmonds, J. Aumentado, and J. D. Teufel,
			Direct observation of deterministic macroscopic entanglement,
			Science {\bf 372}, 622 (2021).
			\bibitem{Lepinay}
			L. Mercier de L\'{e}pinay, C. F. Ockeloen-Korppi, M. J. Woolley, and M. A. Sillanp\"a\"a,
			Quantum-mechanics-free subsystems with mechanical osillators,
			Science {\bf 372}, 625 (2021).
						
			\bibitem{Feynman}
			R. P. Feynman, F. M. Morinigo, and W. G. Wagner, Feynmann Lectures on Gravitation, (Westview Press, Boulder, 1995).
			
			\bibitem{Tabletop}
			D. Carney, P. C. E. Stamp, and J. M. Taylor, 
			Class. and Quant. Grav. {\bf 36} 034001 (2019).
			
			\bibitem{Bose}
			S. Bose, A. Mazumdar, G. W. Morley, H. Ulbricht, M. Toro\v{s}, M. Paternostro, A. A. Geraci, P. F. Barker, M. S. Kim, and G. Milburn,
			Phys. Rev. Lett. $\bm{119}$, 240401 (2017).
			
			\bibitem{MV}
			C. Marletto and V. Vedral,
			Phys. Rev. Lett. $\bm{119}$, 240402 (2017).
			
			\bibitem{Matsumoto}
			N. Matsumoto, S. B. Catan\~{o}-Lopez, M. Sugawara, S. Suzuki, N. Abe, K. Komori, Y. Michimura, Y. Aso, and K. Edamatsu,
			Phys. Rev. Lett. $\bm{122}$, 071101 (2019).
			
			\bibitem{MY}
			N. Matsumoto and N. Yamamoto,  arXiv:2008.10848.

			\bibitem{Meng}
			C. Meng, G. A. Brawley, J. S. Bennett, and W. P. Bowen,
			Phys. Rev. Lett. $\bm{125}$, 043604 (2020).
			
			\bibitem{Blaushi} 
			A. A. Balushi, W. Cong,  and R. B. Mann, 
			Phys. Rev. A {\bf 98} 043811 (2018).
			\bibitem{Miao} 
			H. Miao, D. Martynov, H. Yang, and A. Datta,
			Phys. Rev. A {\bf 101} 063804 (2020).
			\bibitem{Matsumura}
			A. Matsumura and K. Yamamoto,  
			Phys. Rev. D {\bf 102} 106021 (2020).
			\bibitem{Krisnanda}
			T. Krisnanda, G. Y. Tham, M. Paternostro, and T. Paterek,
			npj Quantum Inf. {\bf 6}, 12 (2020).
			
			\bibitem{Datta}
			A. Datta and H. Miao,
			Quantum Sci. Technol. {\bf 6}, 045014 (2021).
			
			\bibitem{Miki2}
			D. Miki, A. Matsumura, and K. Yamamoto, Phys. Rev. D {\bf 105}, 026011 (2022).
			
			\bibitem{Genes}
			C. Genes, D. Vitali, P. Tombesi, S. Gigan, and M. Aspelmeyer
			, Phys. Rev. A {\bf 77} 033804 (2008).
			\bibitem{Vitali}
			D. Vitali et al., Phys. Rev. Lett. {\bf 98} 030405 (2007).
			\bibitem{Miao10}
			H. Miao, S. Danilishin, H. M\"uller-Hbhardt, and Y. Chen, New Journal of Physics, {\bf 12}, 083032 (2010).
	
			\bibitem{Wiener}
			N. Wiener, Extrapolation, Interpolation, and Smoothing of Stationary Time Series ( MIT Press, 1964).

           \bibitem{NYamamoto}
           H. I. Nurdin and N. Yamamoto, {Linear Dynamical Quantum Systems, Analysis, Synthesis, and Control, (Springer, 2017).}
                        
			\bibitem{Schmole}
			J. Schm\'{o}le, M. Dragosits, H. Hepach, and M. Aspelmeyer,
			Classical Quantum Gravity, $\bm{33}$, 125031 (2016).
			
			\bibitem{Lopez}
			S. B. Catan\~{o}-Lopez, J. G. Santiago-Condori, K. Edamatsu, and N. Matsumoto,
			Phys. Rev. Lett. $\bm{124}$, 221102 (2020).
			
			\bibitem{Sugiyama}
			Y. Sugiyama, T. Shichijo, N. Matsumoto, A. Matsumura, D. Miki, and K. Yamamoto, Phys. Rev. A {\bf 107} 033515 (2023).

                      \bibitem{Saulson}
			G. I. Gonzalez and P. R. Saulson, J. Acoust. Soc. Am. {\bf 96}, 207 (1994).

			
			\bibitem{Miki3}
			D. Miki, N. Matsumoto, A. Matsumura, T. Shichijo, Y. Sugiyama, K. Yamamoto, and N. Yamamoto, Phys. Rev. A {\bf 107} 032410 (2023).

   \bibitem{Law}
   C. K. Law, Physical Review A, {\bf 51} 2537 (1995).

   \bibitem{Whittle}
   C. Whittle, et al. Science {\bf 372} 6548, 1333-1336
			(2021).
   
			\bibitem{Giovannetti}
			V. Giovannetti and D. Vitali, Phys. Rev. A {\bf 63}, 023812 (2001).
		
\bibitem{Grover}
R. G. Brown, P. Y. C, Hwang, 
{\it Introduction to Random Signals and Applied Kalman Filtering} (1996, New York: John Wiley \& Sons)

		\end{thebibliography}
	\end{document}